\documentclass[12pt,letter]{article}
\usepackage{amsthm,amssymb, color, epsfig, graphics}
\usepackage{xcolor}
\usepackage{graphicx}
\usepackage[intlimits]{amsmath}
\usepackage{latexsym}

\theoremstyle{definition}

\newtheorem{example}{Example}
\newtheorem{examplemod}{Example}

\begin{document}

%  Headings
%
%\renewcommand{\evenhead}{M. B\l aszak and A. Sergyeyev}
%\renewcommand{\oddhead}{Contact Lax pairs and associated (3+1)-dimensional integrable
%systems}

%  Titlepage
%
%\thispagestyle{empty}

\title{Contact Lax pairs and associated (3+1)-dimensional integrable
dispersionless systems}

\author{Maciej B\l aszak~$^a$ and Artur Sergyeyev~$^b$\\
$^a$ Faculty of Physics, Division of Mathematical Physics,\\ A. Mickiewicz
University\\
Umultowska 85, 61-614 Pozna\'{n}, Poland\\
E-mail \texttt{blaszakm@amu.edu.pl}\\[10pt]
$^b$ Mathematical Institute, Silesian University in Opava,\\
Na Rybn\'\i{}\v{c}ku 1, 74601 Opava, Czech Republic\\
E-mail \texttt{artur.sergyeyev@math.slu.cz}}

\maketitle

\begin{abstract}
\noindent We review the recent approach to the construction of
(3+1)-dimensional integrable dispersionless partial differential systems
based on their contact Lax pairs and the related $R$-matrix theory for the
Lie algebra of functions with respect to the contact bracket. We discuss
various kinds of Lax representations for such systems, in particular, linear
nonisospectral contact Lax pairs and nonlinear contact Lax pairs as well as
the relations among the two. Finally, we present a large number of examples
with finite and infinite number of dependent variables, as well as the
reductions of these examples to lower-dimensional integrable dispersionless
systems.
\end{abstract}

\section{ Introduction}

Integrable systems play an important role in modern mathematics and
theoretical and mathematical physics, cf.\ e.g.\ \cite{d-book, o}, and, since
according to general relativity our spacetime is four-dimensional,
integrable systems in four independent variables ((3+1)D for short; likewise
($n$+1)D is shorthand for $n+1$ independent variables) are particularly
interesting. For a long time it appeared that such systems were very
difficult to find but in a recent paper by one of us \cite{aslmp} a novel
systematic and effective construction for a large new class of integrable
(3+1)D systems was introduced. This construction uses Lax pairs of a new
kind related to contact geometry. Moreover, later in \cite{bls} it was shown
that the systems from this class are amenable to an appropriate extension of
the $R$-matrix approach which paved the way to constructing the associated
integrable hierarchies.

The overwhelming majority of integrable partial differential systems in four
or more independent variables known to date, cf.\ e.g.\ \cite%
{d-book,fkk,ms,ms2} and references therein, including the celebrated
%the ones most relevant for physics like
(anti-)self-dual Yang--Mills equations and (anti-)self-dual vacuum Einstein
equations with vanishing cosmological constant, can be written as
homogeneous first-order quasilinear, i.e., \emph{dispersionless}, also known
as \emph{hydrodynamic-type}, systems, cf.\ e.g.\ \cite{dn,d-book,fkk,z2}
and the discussion below for details on the latter.

Integrable (3+1)D systems from the class introduced in \cite{aslmp} and
further studied in \cite{bls, as-nd, asalp} also are dispersionless, and it
is interesting to note that this class appears to be entirely new: it does
not seem to include any of the previously known examples of integrable
dispersionless (3+1)D systems with nonisospectral Lax pairs, e.g.\ those
from \cite{d, d2, fkk, km}.

In the present paper we review the results from \cite{aslmp, bls} and
provide some novel examples of integrable (3+1)D systems using the approach
from these papers.

The rest of the text is organized as follows. After a brief review of (3+1)D
dispersionless systems and their nonisospectral Lax pairs in general in
Section~\ref{se2} we proceed with recalling the properties of linear and
nonlinear Lax pairs in (1+1)D and (2+1)D in Section~\ref{se3}. In Section~%
\ref{se4} we review, following \cite{aslmp}, the construction of linear and
nonlinear contact Lax pairs and the associated integrable (3+1)D systems and
illustrate it by several examples. Finally, in Section~\ref{se5} we survey,
following \cite{bls}, a version of the $R$-matrix formalism adapted to this
setting and again give a number of examples to illustrate it.\looseness=-1

\section{Isospectral versus nonisospectral Lax pairs}

\label{se2}

Dispersionless systems in four independent variables $x,y,z,t$ by definition
can be written in general form
\begin{equation}  \label{dis}
A_0(\boldsymbol{u})\boldsymbol{u}_t+A_1(\boldsymbol{u})\boldsymbol{u}_x+A_2(%
\boldsymbol{u})\boldsymbol{u}_y+A_3(\boldsymbol{u})\boldsymbol{u}_z=0
\end{equation}
where $\boldsymbol{u}=(u_1,\dots,u_N)^T$ is an $N$-component vector of
unknown functions and $A_i$ are $M\times N$ matrices, $M\geq N$.

Integrable systems of the form (\ref{dis}) typically have scalar Lax
%first-order
pairs of general form
\begin{equation}
\begin{array}{rcl}
\chi _{y} & = & K_{1}(p,\boldsymbol{u})\chi _{x}+K_{2}(p,\boldsymbol{u})\chi
_{z}+K_{3}(p,\boldsymbol{u})\chi _{p}, \\
\chi _{t} & = & L_{1}(p,\boldsymbol{u})\chi _{x}+L_{2}(p,\boldsymbol{u})\chi
_{z}+L_{3}(p,\boldsymbol{u})\chi _{p},%
\end{array}
\label{nil}
\end{equation}%
where $\chi =\chi (x,y,z,t,p)$ and $p$ is the (variable) spectral parameter,
cf.\ e.g.\ \cite{bzm,z,aslmp} and references therein; we stress that $%
\boldsymbol{u}_{p}=0$.

In general, if at least one of the quantities $K_{3}$ or $L_{3}$ is nonzero,
these Lax pairs are nonisospectral as they involve $\chi _{p}$. The same
terminology is applied in the lower-dimension case, when e.g.\ the
dependence on $z$ is dropped. The isospectral case when both $K_{3}$ and $%
L_{3}$ are identically zero is substantially different from the
nonisospectral one. In particular, it is conjectured \cite{fp} that
integrable systems with isospectral Lax pairs (\ref{nil}) are linearly
degenerate while those with nonisospectral Lax pairs (\ref{nil}) are not,
which leads to significant differences in qualitative behavior of solutions:
according to a conjecture of Majda \cite{majda}, in linearly degenerate
systems no shock formation for smooth initial data occurs, see also the
discussion in \cite{fkk}. Many examples of integrable dispersionless (3+1)D
systems with Lax pairs (\ref{nil}) in the isospectral case can be found
e.g.\ in \cite{km, mos, as, as-ro} and references therein. \looseness=-1

On the other hand, it appears that, among dispersionless systems, only
linearly degenerate systems admit recursion operators being B\"acklund
auto-transformations of linearized versions of these systems, cf.\ e.g.\
\cite{mar} and references therein for general introduction to the recursion
operators of this kind, and \cite{mas, mos, as, as-ro} and references
therein for such operators in the context of dispersionless systems. The
theory of recursion operators for integrable dispersionless systems with
nonisospectral Lax pairs (\ref{nil}), if any exists, should be significantly
different both from that of the recursion operators as auto-B\"acklund
transformations of linearized versions of systems under study and from that
of bilocal recursion operators, see e.g.\ \cite{af} and references therein
for the latter.

Finally, in the case of nonisospectral Lax pairs (\ref{nil}) integrability
of associated nonlinear systems is intimately related to the geometry of
characteristic varieties of the latter \cite{fkr, ck}. On the other hand,
for large classes of (1+1)D and (2+1)D dispersionless integrable systems
their nonlinear Lax representations are related to symplectic geometry, see
e.g.\ \cite{Li,blpla,bsz1,bsz,fkk,fkr,fmn,os,aslmp,z} and references
therein, although there are some exceptions, cf.\ e.g.\ \cite{ms,bsz4} and
references therein. As a consequence of this, in the (1+1)D case the systems
under study can be written in the form of the Lax equations which take the form of Hamiltonian dynamics on some Poisson algebras. For the (2+1)D case,
the systems under study can be written as zero-curvature-type equations on
certain Poisson algebras, i.e., as Frobenius integrability conditions for
some pseudopotentials or, equivalently, for Hamiltonian functions from the
Poisson algebra under study. Moreover, thanks to some features of symplectic
geometry, the original nonlinear Lax representations in (1+1)D and (2+1)D
imply linear nonisospectral Lax representations written in terms of
Hamiltonian vector fields of the form (\ref{nil}), as discussed in the next
section.\looseness=-1

In view of the wealth of integrable (2+1)D dispersionless systems it is
natural to look %The search
for new multidimensional integrable systems which are dispersionless, and it
is indeed possible to construct in a systematic fashion such new (3+1)D
systems using \emph{contact} geometry instead of symplectic one in a way
proposed in \cite{aslmp}, and we review this construction below. In
particular, we will show how, using this construction, one obtains a novel
class of nonisospectral Lax pairs together with the associated
zero-curvature-type equations in the framework of Jacobi algebras, i.e., as
Frobenius integrability conditions for contact Hamiltonian functions from
such an algebra.

In what follows we will be interested in the class of dispersionless systems
possessing nonisospectral Lax representations.

\section{Lax representations for dispersionless systems\newline
in (1+1)D and (2+1)D}

\label{se3}

\subsection{Nonlinear Lax pairs in (1+1)D and (2+1)D}

Dispersionless systems in (2+1)D have the form (\ref{dis}) with $A_{3}=0$
and $\boldsymbol{u}_{z}=0$, and these in (1+1)D have the form (\ref{dis})
with $A_{3}=A_{2}=0$ and $\boldsymbol{u}_{z}=\boldsymbol{u}_{y}=0$. For the
overwhelming majority of integrable systems of this kind, see e.g.\ \cite%
{d-book,ms2,z}, there exists a pseudopotential $\psi$ such that the systems
under study can be written as an appropriate
%commutativity conditions for pairs of vector fields of the form
%\begin{equation}\label{comco3d}
%[\partial_y-\mathcal{X}_f,\partial_t-\mathcal{X}_g]=0
%\end{equation}
%, z, fkh, os, ms},
%or, equivalently, as
compatibility condition for a nonlinear (with respect to $\psi$) Lax pair.
%Actually, in 1+1 dimensions the nonlinear
%with respect to pseudopotential $\psi (x,t)$
The said nonlinear Lax pair takes the form (cf.\ e.g.\ \cite{kri2})
\begin{equation}
E=\mathcal{L}(\psi _{x},\boldsymbol{u}),\quad \psi _{t}=\mathcal{B}(\psi
_{x},\boldsymbol{u}),  \label{nLax2d}
\end{equation}%
where $E$ is an arbitrary constant playing the role reminiscent of that of a
spectral parameter for the linear Lax pairs, while in (2+1)D the nonlinear
%with respect to pseudopotential $\psi(x,y,t)$
Lax pair takes the form \cite{z} (cf.\ also e.g.\ \cite{fkr, fmn, aslmp} and
references therein)
\begin{equation}
\psi _{y}=\mathcal{L}(\psi _{x},\boldsymbol{u}),\quad \psi _{t}=\mathcal{B}%
(\psi _{x},\boldsymbol{u}).  \label{nLax3d}
\end{equation}%
The compatibility relations for a Lax pair, which are necessary and
sufficient conditions for the existence of a pseudopotential $\psi$, are
equivalent to a system of PDEs for the vector $\boldsymbol{u}$ of dependent
variables.

Let us illustrate this idea by a simple example.

\begin{example}
\label{Ex1} \emph{\ Let $\boldsymbol{u}=(v_{1},v_{2},u_{0},u_{1})^{T}$ and
take
\begin{equation}
\mathcal{L}(\psi _{x},\boldsymbol{u})=\psi _{x}+u_{0}+u_{1}\psi
_{x}^{-1},\quad \mathcal{B}(\psi _{x},\boldsymbol{u})=v_{1}\psi
_{x}+v_{2}\psi _{x}^{2}.  \label{e1}
\end{equation}%
Compatibility of (\ref{e1}) gives
\begin{equation*}
0=\frac{d\mathcal{L}}{dx}=\psi _{xx}+(u_{0})_{x}+(u_{1})_{x}\psi
_{x}^{-1}-u_{1}\psi _{xx}\psi _{x}^{-2}\Rightarrow u_{1}\psi _{xx}\psi
_{x}^{-2}=\psi _{xx}+(u_{0})_{x}+(u_{1})_{x}\psi _{x}^{-1}
\end{equation*}%
and
\begin{align*}
0& =\frac{d\mathcal{L}}{dt}=\psi _{xt}+(u_{0})_{t}+(u_{1})_{t}\psi
_{x}^{-1}-u_{1}\psi _{xt}\psi _{x}^{-2} \\
& =\frac{d\mathcal{B}}{dx}+(u_{0})_{t}+(u_{1})_{t}\psi _{x}^{-1}-u_{1}\frac{d%
\mathcal{B}}{dx}\psi _{x}^{-2} \\
& =(v_{2})_{x}\psi _{x}^{2}+\left[ (v_{1})_{x}-2v_{2}(u_{0})_{x}\right] \psi
_{x}+\left[ (u_{0})_{t}-2v_{2}(u_{1})_{x}-u_{1}(v_{2})_{x}-v_{1}(u_{0})_{x}%
\right] \\
& \qquad +\left[ (u_{1})_{t}-u_{1}(v_{1})_{x}-v_{1}(u_{1})_{x}\right] \psi
_{x}^{-1}.
\end{align*}%
} \emph{Thus, equating to zero the coefficients at the powers of $\psi _{x}$
in the above equation we obtain the following system:
\begin{equation}
\begin{array}{rcl}
(v_{2})_{x} & = & 0, \\
(v_{1})_{x} & = & 2v_{2}(u_{0})_{x}, \\
(u_{0})_{t} & = & 2v_{2}(u_{1})_{x}+u_{1}(v_{2})_{x}+v_{1}(u_{0})_{x}, \\
(u_{1})_{t} & = & u_{1}(v_{1})_{x}+v_{1}(u_{1})_{x}.%
\end{array}
\label{ex1-sys2d}
\end{equation}%
} \emph{\ In particular, if we put $v_{2}=\mathrm{const}=\frac{1}{2}$ and $%
v_{1}=u_{0}$, we arrive at a two-component dispersionless system in
1+1 dimensions
\begin{equation}
\begin{array}{rcl}
(u_{0})_{t} & = & (u_{1})_{x}+u_{0}(u_{0})_{x}, \\
(u_{1})_{t} & = & u_{1}(u_{0})_{x}+u_{0}(u_{1})_{x}.%
\end{array}
\label{e3}
\end{equation}%
}

\emph{Now turn to the (2+1)D Lax pair (\ref{nLax3d}) with (\ref{e1}). Then
we have
\begin{equation}
\psi _{yt}=\psi _{xt}+(u_{0})_{t}+(u_{1})_{t}\psi _{x}^{-1}-u_{1}\psi
_{xt}\psi _{x}^{-2},  \label{e4}
\end{equation}%
\begin{equation}
\psi _{ty}=(v_{1})_{y}\psi _{x}+v_{1}\psi _{xy}+(v_{2})_{y}\psi
_{x}^{2}+2v_{2}\psi _{x}\psi _{xy}.  \label{e5}
\end{equation}
} \emph{The compatibility of (\ref{nLax3d}) results in
\begin{equation*}
\begin{array}{rcl}
0 & = & \psi _{yt}-\psi _{ty}=\left[
(u_{0})_{t}-2v_{2}(u_{1})_{x}-u_{1}(v_{2})_{x}-u_{1}(v_{2})_{x}-v_{1}(u_{0})_{x}%
\right] \\[2mm]
&  & +\left[ (v_{2})_{x}-(v_{2})_{y}\right] \psi _{x}^{2}+\left[
(v_{1})_{x}-(v_{1})_{y}-2v_{2}(u_{0})_{x}\right] \psi _{x} \\[2mm]
&  & +\left[ (u_{1})_{t}-u_{1}(v_{1})_{x}-v_{1}(u_{1})_{x}\right] \psi
_{x}^{-1}.%
\end{array}%
\end{equation*}%
} \emph{Equating to zero the coefficients at the powers of $\psi_x$ yields
the system
\begin{equation}  \label{ex1-sys3d}
\begin{array}{rcl}
(v_2)_y & = & (v_2)_x, \\
(v_1)_y & = & (v_1)_x-2 v_2 (u_0)_x, \\
(u_{0})_{t} & = & 2v_{2}(u_{1})_{x}+u_{1}(v_{2})_{x}+v_{1}(u_{0})_{x}, \\
(u_{1})_{t} & = & u_{1}(v_{1})_{x}+v_{1}(u_{1})_{x}.%
\end{array}%
\end{equation}
} \emph{If we put $v_{2}=\mathrm{const}=\frac{1}{2}$, we arrive at a
three-component dispersionless system in 2+1 dimensions:
\begin{align}
(u_{0})_{t}&
=(u_{1})_{x}+v_{1}(u_{0})_{x}=(u_{1})_{x}+v_{1}(v_{1})_{x}+v_{1}(v_{1})_{y},
\notag \\
(u_{1})_{t}& =u_{1}(v_{1})_{x}+v_{1}(u_{1})_{x},  \label{e8} \\
(v_{1})_{y}& =(u_{0})_{x}-(v_{1})_{x}.  \notag
\end{align}%
}
\end{example}

\subsection{Basics of Poisson geometry}

Now we shall restate the compatibility conditions for Lax pairs (\ref{nLax2d}) and (\ref{nLax3d}) using the language of symplectic geometry but first we
briefly recall the setting of the latter.\looseness=-1

Namely, consider an even-dimensional ($\dim M=2n$) symplectic manifold $%
(M,\omega)$, where $\omega$ is a closed ($d\omega=0$) differential two-form
which is nondegenerate, i.e., such that the $n$th exterior power of $\omega$
does not vanish anywhere on $M$.

Then for an arbitrary smooth function $H$ on $M$ there exists a unique
vector field $\mathfrak{X}_{H}$ (the Hamiltonian vector field) defined by
\begin{equation}
i_{\mathfrak{X}_{H}}\omega =dH\Longleftrightarrow \mathfrak{X}_{H}=\mathcal{P%
}dH,\quad  \label{hx}
\end{equation}%
where $i_{\mathfrak{X}_{H}}\omega $ is the interior product of vector field $%
\mathfrak{X}_{H}$ with $\omega $, and $\mathcal{P}$ is the associated
implectic bivector, i.e., a nondegenerate Poisson bivector; recall that a
bivector is a skew-symmetric twice contravariant tensor field. Then $%
\mathfrak{X}_{H}$ is referred to as a Hamiltonian vector field with the
Hamiltonian $H$.

Note that a symplectic manifold is a particular case of the more general
Poisson manifold. A Poisson manifold is a pair $(M,\mathcal{P})$ where $%
\mathcal{P}$ is a bivector (i.e., a contravariant rank two skew-symmetric
tensor field) satisfying the following identity:
\begin{equation}
\lbrack \mathcal{P},\mathcal{P}]_{S}=0,  \label{sn}
\end{equation}%
where $[\cdot ,\cdot ]_{S}$ is the Schouten bracket, cf.\ e.g.\ \cite{vai,
Schouten1996}.

The Poisson structure $\mathcal{P}$ induces a bilinear map
%  through a particular Lie bracket
\begin{equation*}
\{\cdot ,\cdot \}_{\mathcal{P}}:\mathcal{F}(M)\times \mathcal{F}%
(M)\longrightarrow \mathcal{F}(M),
\end{equation*}%
in the associative algebra $\mathcal{F}(M)$ of smooth functions on $M$ given
by
\begin{equation}
\{F,G\}_{\mathcal{P}}:=\mathcal{P}(dF,dG),  \label{7.5a}
\end{equation}%
which endows $\mathcal{F}(M)$ with the Lie algebra structure and also
satisfies the Leibniz rule, i.e., the bracket is also a derivation with  respect to
multiplication in the algebra of functions. Such a bracket is called a \emph{%
Poisson bracket}.

It is readily checked that once (\ref{sn}) holds we indeed have

\begin{enumerate}
\item $\{F,G\}_{\mathcal{P}}=-\{G,F\}_{\mathcal{P}},$ \quad (antisymmetry),

\item $\{F,GH\}_{\mathcal{P}}=\{F,G\}_{\mathcal{P}}H+G\{F,H\}_{\mathcal{P}},$
\quad (the Leibniz rule),

\item $\{F,\{H,G\}_{\mathcal{P}}\}_{\mathcal{P}}+\{H,\{G,F\}_{\mathcal{P}%
}\}_{\mathcal{P}}+\{G,\{F,H\}_{\mathcal{P}}\}_{\mathcal{P}}=0,$ \quad (the
Jacobi identity).
\end{enumerate}

For a $2n$-dimensional symplectic manifold, by the Darboux theorem there
exist local coordinates $(x^{i},p_{i})$, $i=1,\dots ,n$, known as the
Darboux coordinates, such that $\omega =d\eta $, where $\eta
=\sum\limits_{i=1}^{n}p_{i}dx^{i}$, and hence\looseness=-1
\begin{equation*}
\omega =d\eta =\sum\limits_{i=1}^{n}dp_{i}\wedge dx^{i},\qquad
P=\sum\limits_{i=1}^{n}\partial _{x^{i}}\wedge \partial _{p_{i}},
\end{equation*}%
\begin{equation}
\mathfrak{X}_{H}=\frac{\partial H}{\partial p_{i}}\frac{\partial }{\partial
x^{i}}-\frac{\partial H}{\partial x^{i}}\frac{\partial }{\partial p_{i}}.
\label{hx1}
\end{equation}%
and
\begin{equation}
\{H,F\}_{P}=\mathfrak{X}_{H}(F)=\frac{\partial H}{\partial p_{i}}\frac{%
\partial F}{\partial x^{i}}-\frac{\partial H}{\partial x^{i}}\frac{\partial F%
}{\partial p_{i}}.  \label{hx2}
\end{equation}

For any $H,F\in \mathcal{F}(M)$ we also have
\begin{equation}
\left[ \mathfrak{X}_{H},\mathfrak{X}_{F}\right] =\mathfrak{X}_{\{H,F\}_{P}},
\label{hx3}
\end{equation}
where $[\cdot ,\cdot ]$ is the usual Lie bracket (commutator) of vector
fields.\looseness=-1

\subsection{Compatibility conditions for Lax pairs via Poisson geometry}

Now let us return to the Lax pair (\ref{nLax2d})
\begin{equation}
E=\mathcal{L}(\psi _{x},\boldsymbol{u})\quad \psi _{t}=\mathcal{B}(\psi _{x},%
\boldsymbol{u})  \label{sg1}
\end{equation}%
for the (1+1)D case when $\boldsymbol{u}=\boldsymbol{u}(x,t)$.

We have
\begin{equation}
0=\frac{d\mathcal{L}}{dx}=\frac{\partial \mathcal{L}}{\partial x}+\frac{%
\partial \mathcal{L}}{\partial \psi _{x}}\psi _{xx}\Longrightarrow \psi _{xx}=-\left(
\frac{\partial \mathcal{L}}{\partial \psi _{x}}\right) ^{-1}\frac{\partial \mathcal{L}%
}{\partial x}  \label{sg2}
\end{equation}%
and so
\begin{align}
0& =\frac{d\mathcal{L}}{dt}=\frac{\partial \mathcal{L}}{\partial t}+\frac{%
\partial \mathcal{L}}{\partial \psi _{x}}\psi _{xt}=\frac{\partial \mathcal{L%
}}{\partial t}+\frac{\partial \mathcal{L}}{\partial \psi _{x}}\frac{\partial
\mathcal{B}}{dx}  \notag \\
& =\frac{\partial \mathcal{L}}{\partial t}+\frac{\partial \mathcal{L}}{%
\partial \psi _{x}}\left( \frac{\partial \mathcal{B}}{\partial x}+\frac{%
\partial \mathcal{B}}{\partial \psi _{x}}\psi _{xx}\right) \overset{(\ref%
{sg2})}{=}\frac{\partial \mathcal{L}}{\partial t}+\frac{\partial \mathcal{L}%
}{\partial \psi _{x}}\frac{\partial \mathcal{B}}{\partial x}-\frac{\partial
\mathcal{L}}{\partial x}\frac{\partial \mathcal{B}}{\partial \psi _{x}}.
\label{sg3}
\end{align}

Thus, the compatibility condition for Lax pair (\ref{nLax2d}) is
equivalently expressed via the so-called Lax equation
\begin{equation}
L_{t}=\{B,L\}_{P},  \label{sg4}
\end{equation}
for a pair of functions $L=\mathcal{L}(p,\boldsymbol{u})$, $B=\mathcal{B}(p,%
\boldsymbol{u})$, %\begin{equation*}
%L=\mathcal{L}(p,\boldsymbol{u}),\ B=\mathcal{B}(p,\boldsymbol{u})
%\end{equation*}
where now $P=\partial_x\wedge \partial_p$ is a Poisson bivector associated
to the symplectic two-form $dp \wedge dx$ on a two-dimensional symplectic
manifold with global Darboux coordinates $(x,p)$. Here $p$ is an additional
independent variable, which in the context of linear Lax pairs will be
identified as a \emph{variable spectral parameter}, see next subsection.

Now turn to the nonlinear Lax pair (\ref{nLax3d})
\begin{equation}
\ \psi_{y}=\mathcal{L}(\psi_{x},\boldsymbol{u}) \quad \psi_{t}=\mathcal{B}%
(\psi _{x},\boldsymbol{u})  \label{sg5}
\end{equation}%
for the (2+1)-dimensional case when $\boldsymbol{u}=\boldsymbol{u}(x,y,t)$.

We have
\begin{equation}
\psi _{yt}=\frac{\partial \mathcal{L}}{\partial t}+\frac{\partial \mathcal{L}%
}{\partial \psi _{x}}\psi _{xt},\quad \psi _{ty}=\frac{\partial \mathcal{B}}{%
\partial y}+\frac{\partial \mathcal{B}}{\partial \psi _{x}}\psi _{xy},
\label{sg6}
\end{equation}%
\begin{equation}
\psi_{tx}=\psi _{xt}=\frac{\partial \mathcal{B}}{\partial x}+\frac{\partial
\mathcal{B}}{\partial \psi _{x}}\psi _{xx},\quad \psi _{yx}=\psi _{xy}=\frac{%
\partial \mathcal{L}}{\partial x}+\frac{\partial \mathcal{L}}{\partial \psi
_{x}}\psi _{xx},  \label{sg7}
\end{equation}%
and thus,
\begin{equation}
0=\psi _{yt}-\psi _{ty}\overset{(\ref{sg6}),(\ref{sg7})}{=}\mathcal{L}_{t}-%
\mathcal{B}_{y}+\frac{\partial \mathcal{L}}{\partial \psi_{x}}\frac{\partial
\mathcal{B}}{\partial x}-\frac{\partial \mathcal{L}}{\partial x}\frac{%
\partial \mathcal{B}}{\partial \psi _{x}}.  \label{sg8}
\end{equation}

The compatibility condition for the Lax pair (\ref{nLax3d}) can be now
written as the so-called zero-curvature-type equation of the form \cite%
{bsz,blpla}
\begin{equation}
L_{t}-B_{y}+\{L,B\}_{P}=0,  \label{sg9}
\end{equation}
for a pair of Lax functions $L=\mathcal{L}(p,\boldsymbol{u})$, $B=\mathcal{B}%
(p,\boldsymbol{u})$.

For an illustration of this alternative form of the compatibility conditions
for our nonlinear Lax pairs let us return to our example. % \ref{Ex1}.

\begin{examplemod}
\label{Ex2} \emph{Let $L(p,\boldsymbol{u})=p+u_{0}+u_{1}p^{-1}$, $B(p,%
\boldsymbol{u})=v_{1}p+v_{2}p^{2}$, where $\boldsymbol{u}%
=(v_{1},v_{2},u_{0},u_{1})^{T}$ and $\boldsymbol{u}=\boldsymbol{u}(x,t)$.
Then (\ref{sg4}) gives
\begin{align*}
0& =L_{t}-\{B,L\}_{P} \\
& =(v_{2})_{x}p^{2}+\left[ (v_{1})_{x}-2v_{2}(u_{0})_{x}\right] p+\left[
(u_{0})_{t}-2v_{2}(u_{1})_{x}-v_{1}(u_{0})_{x}-u_{1}(v_{2})_{x}\right] \\
& +\left[ (u_{1})_{t}-v_{1}(u_{1})_{x}-u_{1}(v_{1})_{x}\right] p^{-1},
\end{align*}%
and equating to zero the coefficients at the powers of $p$, we again obtain
the system (\ref{ex1-sys2d}), where we can put $v_{2}=\mathrm{const}=\frac{1%
}{2}$ and $v_{1}=u_{0}$, and then again arrive at the two-component
dispersionless system (\ref{e3}).}\looseness=-1

\emph{On the other hand, in the (2+1)D case, when $\boldsymbol{u}=%
\boldsymbol{u}(x,y,t)$, the zero-curvature-type equation (\ref{sg9}) gives\looseness=-1
\begin{align*}
0& =L_{t}-B_{y}+\{L,B\}_{P} \\
& =\left[ (v_{2})_{x}-(v_{2})_{y}\right] p^{2}+\left[
(v_{1})_{x}-2v_{2}(u_{0})_{x}-(v_{1})_{y}\right] p \\
&+\left[ (u_{0})_{t}-2v_{2}(u_{1})_{x}-v_{1}(u_{0})_{x}-u_{1}(v_{2})_{x}%
\right] +\left[ (u_{1})_{t}-v_{1}(u_{1})_{x}-u_{1}(v_{1})_{x}\right] p^{-1}.
\end{align*}%
Again, equating to zero the coefficients at the powers of $p$ reproduces the
system (\ref{ex1-sys3d}), and we can put $v_{2}=\mathrm{const}=\frac{1}{2}$
and recover the system (\ref{e8}).\looseness=-1}
\end{examplemod}

\subsection{Linear nonisospectral Lax pairs in (1+1)D and (2+1)D}

The relation (\ref{hx3}) among the Poisson algebra of functions on $M$ and the Lie algebra of Hamiltonian vector fields gives rise to alternative linear
nonisospectral Lax pairs written in terms of Hamiltonian vector fields.

In the (1+1)D case such a linear Lax pair takes the form
\begin{equation}
\mathfrak{X}_{L}(\phi )=\{L,\phi \}_{P}=0,\quad \phi _{t}=\mathfrak{X}%
_{B}(\phi )=\{B,\phi \}_{P},  \label{lL1}
\end{equation}%
where $\phi =\phi(x,t,p)$, and in the (2+1)-dimensional case the form
\begin{equation}
\phi_{y}=\mathfrak{X}_{L}(\phi )=\{L,\phi \}_{P},\quad \phi_{t}=\mathfrak{X}%
_{B}(\phi )=\{B,\phi \}_{P},  \label{lL2}
\end{equation}%
where now $\phi =\phi(x,y,t,p)$.

Here $p$ is an additional independent variable known as the \emph{variable
spectral parameter}, cf.\ e.g.\ \cite{bsz1,bsz2,bzm,d-book,ms} for details;
recall that $\boldsymbol{u}_p\equiv 0$ by assumption.

Since the Hamiltonian vector field with a constant Hamiltonian is
identically zero, the Lax equation (\ref{sg4}) implies the compatibility of (%
\ref{lL1}), and the zero-curvature-type equation (\ref{sg9}) implies the
compatibility of (\ref{lL2}), but not vice versa.\looseness=-1
%the other way around.

Indeed, the compatibility condition for (\ref{lL1}) reads
\begin{equation*}
\left[ \partial _{t}-\mathfrak{X}_{B},\mathfrak{X}_{L}\right] (\phi )=0
\end{equation*}%
\begin{equation*}
\Updownarrow (\ref{hx3})
\end{equation*}%
\begin{equation*}
\mathfrak{X}_{L_{t}-\{B,L\}_{P}}(\phi )=\{L_{t}-\{B,L\}_{P},\phi \}=0,
\end{equation*}%
while the compatibility condition for (\ref{lL2}) takes the form
\begin{equation*}
\left[ \partial _{t}-\mathfrak{X}_{B},\partial _{y}-\mathfrak{X}_{L}\right]
(\phi )=0
\end{equation*}%
\begin{equation*}
\Updownarrow (\ref{hx3})
\end{equation*}%
\begin{equation*}
\mathfrak{X}_{L_{t}-B_{y}+\{L,B\}_{P}}(\phi )=\{L_{t}-B_{y}+\{L,B\}_{P},\phi
\}=0.
\end{equation*}

For an explicit illustration of this we return to our example.

\begin{examplemod}
\label{exa} \emph{\ Again let $\boldsymbol{u}=(v_{1},v_{2},u_{0},u_{1})^{T}$
and
\begin{equation}
L(p,\boldsymbol{u})=p+u_{0}+u_{1}p^{-1},\quad B(p,\boldsymbol{u}%
)=v_{1}p+v_{2}p^{2}.  \label{lab}
\end{equation}%
Then in the (1+1)D case, when $\boldsymbol{u}=\boldsymbol{u}(x,t)$, the Lax
pair (\ref{lL1}) reads
\begin{equation}
\begin{array}{l}
(1-u_{1}/p^{2})\phi _{x}-((u_{0})_{x}+(u_{1})_{x}/p)\phi _{p}=0, \\[2mm]
\phi _{t}=(v_{1}+2pv_{2})\phi _{x}-(p(v_{1})_{x}+p^{2}(v_{2})_{x})\phi _{p},%
\end{array}
\label{linlax-ex1-2d}
\end{equation}%
which can be equivalently written as
\begin{equation}
\hspace*{-3mm}
\begin{array}{rcl}
\phi _{x} & = & \displaystyle\frac{p}{p^{2}-u_{1}}\left[
(u_{1})_{x}+p(u_{0})_{x}\right] \phi _{p}, \\[4mm]
\phi _{t} & = & \displaystyle\frac{p}{p^{2}-u_{1}}((v_{1}+2pv_{2})\left[
(u_{1})_{x}+p(u_{0})_{x}\right] -(p^{2}-u_{1})\left[ (v_{1})_{x}+(v_{2})_{x}p%
\right] )\phi _{p}.%
\end{array}%
\hspace{-2mm}  \label{linlax-ex1-2d-v2}
\end{equation}%
}\emph{The compatibility condition for (\ref{linlax-ex1-2d-v2}) is just $%
(\phi _{x})_{t}-(\phi _{t})_{x}=0$ but we cannot reproduce directly (\ref%
{ex1-sys2d}) by equating to zero the coefficients at the powers of $p.$
Instead, we get a set of linear combinations of differential consequences of
the latter.}

\emph{Now turn to the (2+1)D case with the same $L$ and $B$ given by (\ref%
{lab}) but with $\boldsymbol{u}=\boldsymbol{u}(x,y,t)$. The associated
linear nonisospectral Lax pair (\ref{lL2}) takes the form (\ref{nil}), i.e.
\begin{equation}
\begin{array}{rcl}
\phi _{y} & = & (1-u_{1}/p^{2})\phi _{x}-((u_{0})_{x}+(u_{1})_{x}/p)\phi
_{p}, \\[2mm]
\phi _{t} & = & (v_{1}+2pv_{2})\phi _{x}-(p(v_{1})_{x}+p^{2}(v_{2})_{x})\phi
_{p},%
\end{array}
\label{linlax-ex1-3d}
\end{equation}%
and, in complete analogy with the (1+1)D case, it is readily checked that
its compatibility condition $(\phi _{y})_{t}-(\phi _{t})_{y}=0$ holds by
virtue of the zero-curvature-type equation $L_{t}-B_{y}+\{L,B\}_{P}=0$ but
not the other way around.} \emph{Besides, just as in the (1+1)D case above,
equating to zero the coefficients at the powers of $p$ in $%
(\phi_{y})_{t}-(\phi_{t})_{y}=0$, yields a system being a mix of algebraic
and differential consequences of (\ref{ex1-sys3d}).}\looseness=-1
\end{examplemod}

\section{Lax representations for dispersionless systems in (3+1)D}

\label{se4}

\subsection{Nonlinear Lax pairs in (3+1)D}

In \cite{aslmp} the following generalization of the (2+1)D nonlinear Lax
pair (\ref{nLax3d}) to (3+1)D was found:
\begin{equation}
\psi _{y}=\psi _{z}\mathcal{L}\left( \frac{\psi _{x}}{\psi _{z}},\boldsymbol{%
u}\right) ,\quad \psi _{t}=\psi _{z}\mathcal{B}\left( \frac{\psi _{x}}{\psi
_{z}},\boldsymbol{u}\right) ,  \label{nLax4d}
\end{equation}%
where now $\psi =\psi (x,y,z,t)$. The Lax pairs of the form (\ref{nLax4d})
are called \emph{nonlinear contact Lax pairs}.

The above generalization leads to large new classes of integrable (3+1)D
dispersionless systems for suitably chosen $\mathcal{L}$ and $\mathcal{B}$,
e.g.\ rational functions or polynomials in $\psi _{x}/\psi _{z}$ of certain
special form.

The compatibility conditions for the Lax pair (\ref{nLax4d}), which are
necessary and sufficient conditions for the existence of a nontrivial
pseudopotential $\psi $, are equivalent to a system of PDEs for $\boldsymbol{%
u}$ in (3+1)D.\looseness=-1

Let us illustrate this idea again on our simple example.

\begin{examplemod}
\label{Ex3} \emph{Let
\begin{equation}
\hspace*{-7mm}
\begin{array}{rcl}
\psi _{y} & = & \displaystyle\psi _{z}\mathcal{L}\left( \frac{\psi _{x}}{%
\psi _{z}},\boldsymbol{u}\right) =\psi _{z}\left( \frac{\psi_{x}}{\psi _{z}}%
+u_{0}+u_{1}\biggl(\frac{\psi _{x}}{\psi _{z}}\biggr)^{-1}\right) =\psi
_{x}+u_{0}\psi _{z}+u_{1}\frac{\psi _{z}^{2}}{\psi _{x}}, \\[5mm]
\quad\psi _{t} & = & \displaystyle\psi _{z} \mathcal{B}\left(\frac{\psi _{x}%
}{\psi _{z}},\boldsymbol{u}\right) =\psi _{z}\left( v_{1}\frac{\psi _{x}}{%
\psi _{z}}+v_{2}\biggl(\frac{\psi_{x}}{\psi_{z}}\biggr)^{2}\right)
=v_{1}\psi _{x}+v_{2}\frac{\psi _{x}^{2}}{\psi_{z}},%
\end{array}
\label{3d}
\end{equation}%
where $\boldsymbol{u}=(v_{1},v_{2},u_{0},u_{1})^{T}$. } \emph{Then we have
\begin{align}
\psi _{yt}& =\psi _{xt}+(u_{0})_{t}\psi _{z}+u_{0}\psi _{zt}+(u_{1})_{t}%
\frac{\psi _{z}^{2}}{\psi _{x}}+2u_{1}\frac{\psi _{z}\psi _{zt}}{\psi _{x}}%
-u_{1}\frac{\psi _{z}^{2}\psi _{xt}}{\psi _{x}^{2}},  \notag \\
\psi _{ty}& =(v_{1})_{y}\psi _{x}+v_{1}\psi _{xy}+(v_{2})_{y}\frac{\psi
_{x}^{2}}{\psi _{z}}+2v_{2}\frac{\psi _{x}\psi _{xy}}{\psi _{z}^{2}}-v_{2}%
\frac{\psi _{x}^{2}\psi _{zy}}{\psi _{z}^{2}}  \label{3da}
\end{align}%
} \emph{and, the compatibility of (\ref{3da}) results in
\begin{equation}
\begin{array}{rcl}
0 & = & \psi _{yt}-\psi _{ty}\overset{}{=}\left[
(v_{2})_{x}+u_{0}(v_{2})_{z}-(v_{2})_{y}+v_{2}(u_{0})_{z}\right] %
\displaystyle\frac{\psi _{x}^{2}}{\psi _{z}} \\[2mm]
&  & +\left[ (u_{1})_{t}-u_{1}(v_{1})_{x}-v_{1}(u_{1})_{x}\right] %
\displaystyle\frac{\psi _{z}^{2}}{\psi _{x}} \\[3mm]
&  & +\left[
(v_{1})_{x}+u_{0}(v_{1})_{z}+2u_{1}(v_{2})_{z}-(v_{1})_{y}-2v_{2}(u_{0})_{x}+v_{2}(v_{1})_{z}%
\right] \psi _{x} \\[2mm]
&  & +\left[
(u_{0})_{t}-u_{1}(v_{2})_{x}+2u_{1}(v_{1})_{z}-v_{1}(u_{0})_{x}-2v_{2}(u_{1})_{x}%
\right]\psi _{z}%
\end{array}
\label{3dc}
\end{equation}
and we arrive at a four-component dispersionless (3+1)D integrable system
\begin{equation}
\begin{array}{rcl}
(u_{1})_{t} & = & u_{1}(v_{1})_{x}+v_{1}(u_{1})_{x}, \\
(u_{0})_{t} & = &
u_{1}(v_{2})_{x}-2u_{1}(v_{1})_{z}+v_{1}(u_{0})_{x}+2v_{2}(u_{1})_{x}, \\
(v_{1})_{y} & = &
(v_{1})_{x}+u_{0}(v_{1})_{z}+2u_{1}(v_{2})_{z}-2v_{2}(u_{0})_{x}+v_{2}(v_{1})_{z},
\\
(v_{2})_{y} & = & (v_{2})_{x}+u_{0}(v_{2})_{z}+v_{2}(u_{0})_{z}.%
\end{array}
\label{3dd}
\end{equation}%
}

\emph{The fields $u_{0}$ and $u_{1}$ are dynamical variables, which
evolve in time, while the remaining equations can be seen as
nonlocal constraints on $u_{0}$ and $u_{1}$ which define the variables $%
v_{1} $ and $v_{2}$. The same situation takes place in (2+1)D case. In the
(1+1)D case all fields $v_i$ are expressible via the dynamical fields $u_j$.}
\end{examplemod}

\subsection{Basics of contact geometry}

Now let us restate the compatibility conditions for Lax pairs (\ref{nLax4d})
in the language of contact geometry, and to this end we first recall the basics of the latter.

Consider an odd-dimensional ($\dim M=2n+1$) \emph{contact manifold} $(M,\eta)$
with a contact one-form $\eta$ such that $\eta \wedge (d\eta)^{\wedge n}\neq
0$, cf.\ e.g.\ \cite{br} and references therein.

For a given contact form $\eta $ there exists a unique vector field $Y$,
called the \emph{Reeb vector field}, such that
\begin{equation}
i_{Y}d\eta =0,\quad \ i_{Y}\eta =1.  \label{con1}
\end{equation}%
For any function on $M$, there exists a unique vector field $X_{H}$ (the
\emph{contact vector field}) defined by the formula
\begin{equation}
i_{X_{H}}\eta =H,\quad i_{X_{H}}d\eta =dH-i_{Y}dH\cdot \eta
\Longleftrightarrow X_{H}=\mathcal{P}dH+HY,\quad  \label{con2}
\end{equation}%
where $\mathcal{P}$ is the associated bivector.

Contact manifold is a special case of the so-called Jacobi manifold. A \emph{Jacobi manifold} \cite{lic} is a triple $(M,\mathcal{P},Y)$ where $\mathcal{P%
}$ is a bivector
%(a contravariant 2-tensor)
and $Y$ a vector field satisfying the following conditions:
\begin{equation}
\lbrack \mathcal{P},\mathcal{P}]_{S}=2Y\wedge \mathcal{P},\quad \lbrack Y,%
\mathcal{P}]_{S}=0.  \label{con3}
\end{equation}%
The Jacobi structure induces a bilinear map $\{\cdot ,\cdot \}_{J}:\mathcal{F%
}(M)\times \mathcal{F}(M)\longrightarrow \mathcal{F}(M)$ in the associative
algebra $\mathcal{F}(M)$ of smooth functions on $M$ through the \emph{Jacobi
bracket}
\begin{equation}
\{F,G\}_{J}:=\mathcal{P}(dF,dG)+FY(G)-GY(F),  \label{con4}
\end{equation}%
which turns $\mathcal{F}(M)$ into a Lie algebra and satisfies the
generalized Leibniz rule, i.e., we have

\begin{enumerate}
\item $\!\{F,G\}_{J}=-\{G,F\}_{J}$ \ (antisymmetry),

\item $\!\{F,GH\}_{J}\!=\{F,G\}_{J}H+G\{F,H\}_{J}-\{F,1\}_{J}GH$ (the
generalized Leibniz rule),

\item $\!\{F,\{H,G\}_{J}\}_{J}+\{H,\{G,F\}_{J}\}_{J}+\{G,\{F,H\}_{J}\}_{J}=0$
\ (the Jacobi identity).
\end{enumerate}

For a $(2n+1)$-dimensional contact manifold by the Darboux theorem there
exist local coordinates $(x^{i},p_{i},z)$, where $i=1,\dots ,n$, known as
the Darboux coordinates, such that we have
\begin{equation*}
\eta =dz+\sum\limits_{i=1}^{n}p_{i}dx^{i}\Rightarrow d\eta
=\sum\limits_{i=1}^{n}dp_{i}\wedge dx^{i},\quad Y=\partial _{z},\quad
P=\sum\limits_{i=1}^{n}\left( \partial _{x^{i}}\wedge \partial
_{p_{i}}-p_{i}\partial _{z}\wedge \partial_{p_{i}}\right),
\end{equation*}%
\begin{equation}
X_{H}=H\frac{\partial }{\partial z}+\sum\limits_{i=1}^{n}\left( \frac{%
\partial H}{\partial p_{i}}\frac{\partial }{\partial x^{i}}-\frac{\partial H%
}{\partial x^{i}}\frac{\partial }{\partial p_{i}}-p_{i}\left( \frac{\partial
H}{\partial p_{i}}\frac{\partial }{\partial z}-\frac{\partial H}{\partial z}%
\frac{\partial }{\partial p_{i}}\right) \right)  \label{con6}
\end{equation}%
and the contact bracket, the relevant special case of the Jacobi bracket, reads
\begin{equation}
\hspace*{-4mm} \{H,F\}_{C}\!=\!X_{H}(F)-Y(H)F=H\frac{\partial F}{\partial z}%
+\!\sum\limits_{i=1}^{n}\left(\frac{\partial H}{\partial p_{i}}\frac{%
\partial F}{\partial x^{i}}-p_{i}\frac{\partial H}{\partial p_{i}}\frac{%
\partial F}{\partial z}\right)-\left( H\!\leftrightarrow\! F\right)\!.%
\hspace{-1mm}  \label{con7}
\end{equation}%
We also have
\begin{equation}
\left[ X_{H},X_{F}\right] =X_{\{H,F\}_{C}}.  \label{con8}
\end{equation}

\subsection{Zero-curvature-type equations in (3+1)D via the contact bracket}

Now return to the Lax pair (\ref{nLax4d}) with $\boldsymbol{u}=%
\boldsymbol{u}(x,y,z,t)$,
\begin{equation}
\psi _{y}=\psi _{z}\mathcal{L}\biggl(\frac{\psi _{x}}{\psi _{z}},\boldsymbol{%
u}\biggr),\quad \psi _{t}=\psi _{z}\mathcal{B}\biggl(\frac{\psi _{x}}{\psi
_{z}},\boldsymbol{u}\biggr).  \label{c1}
\end{equation}%
Let %\begin{equation*}
$\theta \equiv \psi _{x}/\psi _{z}$. %\end{equation*}%
Then we have
\begin{align}
\psi _{yt}& =\psi _{zt}\mathcal{L}+\psi _{z}\mathcal{L}_{t}+\psi _{xt}%
\mathcal{L}_{\theta }-\psi _{zt}\theta \mathcal{L}_{\theta },  \notag \\
\psi _{ty}& =\psi _{zy}\mathcal{B}+\psi _{z}\mathcal{B}_{y}+\psi _{xy}%
\mathcal{B}_{\theta }-\psi _{zy}\theta \mathcal{B}_{\theta }.  \label{c2}
\end{align}%
Again, the compatibility of (\ref{c1})\ results in
\begin{equation}
\hspace*{-3mm} 0=\psi_{yt}-\psi_{ty}\!=\psi _{z}\!\left[\mathcal{L}_{t}-%
\mathcal{B}_{y}+\mathcal{L}_{\theta }\mathcal{B}_{x}-\mathcal{L}_{x}\mathcal{%
B}_{\theta }-\theta \left( \mathcal{L}_{\theta }\mathcal{B}_{z}-\mathcal{L}%
_{z}\mathcal{B}_{\theta }\right) +\mathcal{LB}_{z}-\mathcal{BL}_{z}\right].%
\hspace{-1mm}  \label{c4}
\end{equation}

Comparing (\ref{c4}) with (\ref{con7}) we observe that compatibility
condition for Lax pair (\ref{nLax4d}) is equivalently given \cite{aslmp} by
the so-called zero-curvature-type equation of the form
\begin{equation}
L_{t}-B_{y}+\{L,B\}_{C}=0,  \label{c5}
\end{equation}
for a pair of Lax functions $L=\mathcal{L}(p,\boldsymbol{u})$, $B=\mathcal{B}(p,\boldsymbol{u})$,
where the contact bracket $\{\cdot,\cdot\}_C$ now is a special case of the
contact bracket (\ref{con7}) for the three-dimensional contact manifold with
the (global) Darboux coordinates $(x,p,z)$, where $p$ is the variable
spectral parameter just as in the lower-dimensional cases.

The contact bracket in this case reads
\begin{equation}
\{H,F\}_{C}=X_{H}(F)-Y(H)F=H\frac{\partial F}{\partial z}+\frac{\partial H}{%
\partial p}\frac{\partial F}{\partial x}-p\frac{\partial H}{\partial p}\frac{%
\partial F}{\partial z}-\left( H\leftrightarrow F\right) .  \label{con7a}
\end{equation}

For the illustration of that alternative Lax representation let us return to
our previous example.

\begin{examplemod}
\label{Ex4} \emph{Let
\begin{equation}
L(p,\boldsymbol{u})=p+u_{0}+u_{1}p^{-1},\quad B(p,\boldsymbol{u})=v_{1}p+v_{2}p^{2},  \label{xx}
\end{equation}%
where $\boldsymbol{u}=(v_{1},v_{2},u_{0},u_{1})^{T}$. Then, for
(3+1)-dimensional case, the contact zero-curvature-type equation (\ref{c5})
reads
\begin{equation*}
\begin{array}{rcl}
0& =&L_{t}-B_{y}+\{L,B\}_{C}=\left[
(v_{2})_{x}-(v_{2})_{y}+u_{0}(v_{2})_{z}+v_{2}(u_{0})_{z}\right] p^{2} \\[2mm]
&&+\left[
(v_{1})_{x}-2v_{2}(u_{0})_{x}-(v_{1})_{y}+2u_{1}(v_{2})_{z}+v_{2}(u_{1})_{z}+u_{0}(v_{1})_{z}%
\right] p \\[2mm]
&&+\left[
(u_{0})_{t}-2v_{2}(u_{1})_{x}-v_{1}(u_{0})_{x}-u_{1}(v_{2})_{x}+2u_{1}(v_{1})_{z}%
\right] \\[2mm]
&&+\left[ (u_{1})_{t}-v_{1}(u_{1})_{x}-u_{1}(v_{1})_{x}\right]
p^{-1}.
\end{array}\end{equation*}%
and we recover the four-component (3+1)-dimensional integrable
dispersionless system (\ref{3dd}).}
\end{examplemod}

\subsection{Linear nonisospectral Lax pairs in (3+1)D}

Using the above results from contact geometry we readily can construct \cite%
{aslmp} two different kinds of linear nonisospectral Lax pairs in (3+1)D
generalizing (\ref{lL2}), that is,\looseness=-1
\begin{equation*}
\phi _{y}=\mathfrak{X}_{L}(\phi)=\{L,\phi\}_{P},\quad \phi _{t}=\mathfrak{X}%
_{B}(\phi)=\{B,\phi \}_{P},
\end{equation*}%
in two different ways.

The first one replaces the Poisson bracket $\{\cdot ,\cdot \}_{P}$ by the
contact bracket (\ref{con7a}) and gives us the Lax pair of the form
\begin{equation}
\phi _{y}=\{L,\phi \}_{C},\quad \phi _{t}=\{B,\phi \}_{C},  \label{lL3a}
\end{equation}%
where now $\phi =\phi (x,y,z,t,p)$.

The second one replaces the Hamiltonian vector fields $\mathfrak{X}_{H}$ by
their contact counterparts $X_{H}$, and we obtain
\begin{equation}
\chi _{y}=X_{L}(\chi ),\quad \chi _{t}=X_{B}(\chi ),  \label{lL3b}
\end{equation}%
where now $\chi =\chi (x,y,z,t,p)$; here we replaced $\phi $ by $\chi $ in
order to distinguish (\ref{lL3a}) from (\ref{lL3b}). The Lax pairs of the
form (\ref{lL3b}) are called \emph{linear contact Lax pairs} \cite{aslmp}. %
\looseness=-1

Recall that in our particular setting we have
\begin{equation}  \label{cvf}
X_H=H\frac{\partial }{\partial z} +\frac{\partial H}{\partial p}\frac{%
\partial }{\partial x}-\frac{\partial H}{\partial x}\frac{\partial }{%
\partial p}-p\left( \frac{\partial H}{\partial p}\frac{\partial }{\partial z}%
-\frac{\partial H}{\partial z}\frac{\partial }{\partial p}\right),\quad Y=%
\frac{\partial}{\partial z}.
\end{equation}

In stark contrast with the (2+1)D case, the two Lax pairs (\ref{lL3a}) and (%
\ref{lL3b}) no longer coincide, since we have
\begin{equation}  \label{cvfcb}
X_{H}(F)=\{H,F\}_{C}+FH_{z}=\{H,F\}_{C}+F Y(H)=\{H,F\}_{C}+F \{1,H\}_C
\end{equation}%
instead of
\begin{equation*}
\mathfrak{X}_{H}(F)=\{H,F\}_{P},
\end{equation*}%
and the behaviour of these Lax pairs is quite different too.

As for (\ref{lL3a}), in complete analogy with the (2+1)D case we readily
find that its compatibility condition, $(\phi _{y})_{t}-(\phi _{t})_{y}=0$,
can be written as %has the form
\begin{equation}
\{L_{t}-B_{y}+\{L,B\}_{C},\phi \}_{C}=0,  \label{cc-4d-1}
\end{equation}%
where $L=\mathcal{L}(p,\boldsymbol{u})$, $B=\mathcal{B}(p,\boldsymbol{u})$,
and thus while the zero-curvature-type equation (\ref{c5}), or equivalently,
the compatibility of nonlinear Lax pair (\ref{nLax4d}), implies
compatibility of the Lax pair (\ref{lL3a}) but the converse is not true.

On the other hand, the situation for (\ref{lL3b}) is very different. We can,
by analogy with the discussion before Example~\ref{exa}, show that the
compatibility condition for (\ref{lL3b}), that is,
\begin{equation*}
\left[ \partial _{t}-X_{B},\partial_{y}-X_{L}\right]=0
\end{equation*}
%\begin{equation*}
%\Updownarrow
is, by virtue of (\ref{con8}), equivalent to the following:
%\end{equation*}%
\begin{equation}  \label{cvfcc}
X_{L_{t}-B_{y}+\{L,B\}_{C}}=0.
\end{equation}
Using the formula (\ref{cvf}) we immediately see that, in contrast with the
(2+1)-dimensional case, (\ref{cvfcc}) implies that
\begin{equation*}
L_{t}-B_{y}+\{L,B\}_{C}=0,
\end{equation*}
i.e., (\ref{cvfcc}) is \emph{equivalent} to (\ref{c5}) rather than being
just a consequence of the latter, as it is the case for (\ref{cc-4d-1}).\looseness=-1

Let us show the explicit form of the above nonisospectral Lax pairs (\ref%
{lL3a}) and (\ref{lL3b}) for our example.

\begin{examplemod}
\emph{Again, let}
\begin{equation}  \label{lnb}
L(p,\boldsymbol{u})=p+u_{0}+u_{1}p^{-1},\quad B(p,\boldsymbol{u})=v_{1}p+v_{2}p^{2},
\end{equation}%
\emph{where $\boldsymbol{u}=(v_{1},v_{2},u_{0},u_{1})^{T}$. Then the
nonisospectral Lax pair (\ref{lL3a}) reads}%
\begin{equation*}
\begin{array}{rcl}
\phi _{y} & = & (1-u_{1}/p^{2})\phi _{x}+(u_{0}+2u_{1}/p)\phi
_{z}+(p(u_{0})_{z}+(u_{1})_{z}-(u_{0})_{x}-(u_{1})_{x}/p)\phi _{p} \\
&  & +(-(u_{0})_{z}-(u_{1})_{z}/p)\phi, \\[2mm]
\phi _{t} & = & (2pv_{2}+v_{1})\phi _{x}-p^{2}v_{2}\phi
_{z}+((v_{2})_{z}p^{3}+((v_{1})_{z}-(v_{2})_{x})p^{2}-(v_{1})_{x}p)\phi _{p}
\\
&  & -p((v_{2})_{z}p-(v_{1})_{z})\phi,%
\end{array}%
\end{equation*}%
\emph{while the nonisospectral linear contact Lax pair (\ref{lL3b}) has the
form (\ref{nil}), i.e.}
\begin{equation}
\hspace*{-5mm}
\begin{array}{rcl}
\chi _{y} & \!\!=\!\! & \displaystyle\biggl(1-\frac{u_{1}}{p^{2}}\!\biggr)%
\chi _{x}+\biggl(u_{0}+\frac{2u_{1}}{p}\!\biggr)\chi _{z}+\biggl(%
p(u_{0})_{z}+(u_{1})_{z}-(u_{0})_{x}-\frac{(u_{1})_{x}}{p}\!\biggr)\chi _{p},
\\[4mm]
\chi _{t} & \!\!=\!\! & (2pv_{2}+v_{1})\chi _{x}-p^{2}v_{2}\chi _{z}+\biggl(%
(v_{2})_{z}p^{3}+((v_{1})_{z}-(v_{2})_{x})p^{2}-(v_{1})_{x}p\biggr)\chi _{p}.%
\end{array}%
\hspace{-2mm}  \label{4d-lax-ex}
\end{equation}%
\emph{Spelling out the compatibility condition for this Lax pair, $(\chi
_{y})_{t}-(\chi _{t})_{y}=0$, and equating to zero the coefficients at $\chi
_{x}$ and $\chi _{p}$ therein, we readily see that, in perfect agreement
with general discussion above, we recover (\ref{c5}) for $L$ and $B$ given
by (\ref{lnb}), and then the system (\ref{3dd}).}

\emph{As for (\ref{lL3a}), it is readily checked that (\ref{c5}) and (\ref%
{lnb}) imply compatibility of (\ref{lL3a}), that is, $(\phi_y)_t-(%
\phi_t)_y=0 $, but not the other way around, i.e., $(\phi_y)_t-(\phi_t)_y=0$
gives us not the system (\ref{3dd}) but merely a mix of differential and
algebraic consequences thereof.}
\end{examplemod}

\section{$\boldsymbol{R}$-matrix approach for dispersionless systems with nonisospectral Lax representations}

\label{se5}

\subsection{General construction}

The $R$-matrix approach addresses two important problems concerning the
dispersionless systems under study. First, it allows for a systematic
construction of consistent Lax pairs $(L,B)$ in order to generate such
systems, and second, it allows for a systematic construction of an infinite
hierarchy of commuting symmetries for a given dispersionless system, proving
integrability of the latter. So, let us start from some basic facts on the $R$
-matrix formalism, see for example \cite{bsz2,sem,se} and references therein.\looseness=-1

Let $\mathfrak{g}$ be an (in general infinite-dimensional) Lie algebra. The Lie bracket $%
[\cdot ,\cdot ]$ defines the adjoint action of $\mathfrak{g}$ on $\mathfrak{g%
}$: $\mathrm{ad}_{a}b=[a,b]$.

Recall that an $R\in \mathrm{End}(\mathfrak{g})$ is called a (classical) $R$%
-matrix if the $R$-bracket
\begin{equation}
[a,b]_{R}:=[Ra,b]+[a,Rb]  \label{2.2}
\end{equation}%
is a new Lie bracket on $\mathfrak{g}$. The skew symmetry of (\ref{2.2}) is
obvious. As for the Jacobi identity for (\ref{2.2}), a sufficient condition
for it to hold is the so-called classical modified Yang--Baxter equation for
$R$,
\begin{equation}
\lbrack Ra,Rb]-R[a,b]_{R}-\alpha \lbrack a,b]=0,\qquad \alpha \in \mathbb{R}.
\label{2.3}
\end{equation}

Let $L_{i}\in \mathfrak{g}$, $i\in \mathbb{N}$. Consider the associated
hierarchies of flows (Lax hierarchies)
\begin{equation}
(L_{n})_{t_{r}}=[RL_{r},L_{n}],\qquad r,n\in \mathbb{N}.  \label{2.5}
\end{equation}%
Suppose that $R$ commutes with all derivatives $\partial _{t_{n}}$, i.e.,
\begin{equation}
(RL)_{t_{n}}=RL_{t_{n}},\quad n\in \mathbb{N},  \label{2.6}
\end{equation}%
and obeys the classical modified Yang--Baxter equation (\ref{2.3}) for $%
\alpha \neq 0$. Moreover, let $L_{i}\in \mathfrak{g}$, $i\in \mathbb{N}$
satisfy (\ref{2.5}). Then the following conditions are equivalent:

\begin{itemize}
\item[i)] the zero-curvature equations%
\begin{equation}
(RL_{r})_{t_{s}}-(RL_{s})_{t_{r}}+[RL_{r},RL_{s}]=0,\quad r,s\in\mathbb{N}
\label{2.7}
\end{equation}
hold;

\item[ii)] all $L_{i}$ commute in $\mathfrak{g}$:
\begin{equation}
[L_{i},L_{j}]=0, \qquad i,j\in\mathbb{N}.  \label{2.4}
\end{equation}
\end{itemize}

Moreover, if one (and hence both) of the above equivalent conditions holds,
then the flows (\ref{2.5}) %pairwise
commute, i.e.,
\begin{equation}
((L_{n})_{t_{r}})_{t_{s}}-((L_{n})_{t_{s}})_{t_{r}}=0,\quad n,r,s\in \mathbb{%
N}  \label{2.8}
\end{equation}
The reader can find the proofs of the above results for example in \cite{bsz2}
or in \cite{bls}.

Now let us present a procedure for extending the systems under study by adding an
extra independent variable. This procedure bears some resemblance to that of
the central extension approach, see e.g.\ \cite{bsz2,se} and references
therein. Namely, we assume that all elements of $\mathfrak{g}$ depend on
an additional independent variable $y$ not involved in the Lie bracket, so
all of the above results remain valid. Consider an $L\in \mathfrak{g}$ and
the associated Lax hierarchy defined by
\begin{equation}
L_{t_{r}}=[RL_{r},L]+(RL_{r})_{y},\qquad r\in \mathbb{N}.  \label{2.12}
\end{equation}%
Suppose that $L_{i}\in \mathfrak{g}$, $i\in \mathbb{N}$ are such that the
zero-curvature equations (\ref{2.7}) hold for all $r,s\in \mathbb{N}$ and
the $R$-matrix $R$ on $\mathfrak{g}$ satisfies (\ref{2.6}). Then the flows (%
\ref{2.12}) commute, i.e.,
\begin{equation}
(L_{t_{r}})_{t_{s}}-(L_{t_{s}})_{t_{r}}=0,\quad r,s\in \mathbb{N}.
\label{2.13}
\end{equation}

Indeed, using the equations (\ref{2.12}) and the Jacobi identity for the Lie
bracket we obtain
\begin{eqnarray*}
(L_{t_{r}})_{t_{s}}-(L_{t_{s}})_{t_{r}} &=&\left[
(RL_{r})_{t_{s}}-(RL_{s})_{t_{r}}+[RL_{r},RL_{s}],L\right] \\
&&+\left( (RL_{r})_{t_{s}}-(RL_{s})_{t_{r}}+[RL_{r},RL_{s}]\right) _{y} \\
&=&0.
\end{eqnarray*}%
The right-hand side of the above equation vanishes by virtue of the zero
curvature equations (\ref{2.7}).

An important question is whether there exists a systematic procedure for
constructing $R\in \mathrm{End}(\mathfrak{g})$ with the desired properties.
Fortunately, the answer is positive. It is well known (see e.g.\ \cite%
{sem,se,bsz2}) that whenever $\mathfrak{g}$ admits a decomposition into two
Lie subalgebras $\mathfrak{g}_{+}$ and $\mathfrak{g}_{-}$ such that
\begin{equation*}
\mathfrak{g}=\mathfrak{g}_{+}\oplus \mathfrak{g}_{-},\qquad \lbrack
\mathfrak{g}_{\pm },\mathfrak{g}_{\pm }]\subset \mathfrak{g}_{\pm },\qquad
\mathfrak{g}_{+}\cap \mathfrak{g}_{-}=\emptyset ,
\end{equation*}%
the operator
\begin{equation}
R=\frac{1}{2}(\Pi_{+}-\Pi_{-})=\Pi_{+}-\frac{1}{2}  \label{2.14}
\end{equation}%
where $\Pi_{\pm }$ are projectors onto $\mathfrak{g}_{\pm }$, satisfies the
classical modified Yang--Baxter equation (\ref{2.3}) with $\alpha =\frac{1}{4%
}$, i.e., $R$ is a classical $R$-matrix.

Next, we specify the dependence of $L_{j}$ on $y$ via the so-called
Lax--Novikov equations (cf.\ e.g.\  \cite{bsz} and references therein)
\begin{equation}
\lbrack L_{j},L]+(L_{j})_{y}=0,\qquad j\in \mathbb{N}.  \label{2.11}
\end{equation}%
Then, upon applying (\ref{2.4}), (\ref{2.14}) and (\ref{2.11}), after
elementary computations, equations (\ref{2.5}), (\ref{2.7}) and (\ref{2.12})
take the following form:
\begin{equation}
(L_{s})_{t_{r}}=[B_{r},L_{s}],\qquad r,s\in \mathbb{N},  \label{2.15}
\end{equation}%
\begin{equation}
(B_{r})_{t_{s}}-(B_{s})_{t_{r}}+[B_{r},B_{s}]=0,  \label{2.16}
\end{equation}%
\begin{equation}
L_{t_{r}}=[B_{r},L]+(B_{r})_{y},\qquad n,r\in \mathbb{N},  \label{2.17}
\end{equation}%
where $B_{i}=\Pi_{+}L_{i}$.

Obviously, if under the reduction to the case when all quantities are
independent of $y$ we put $L=L_{n}$ for some $n\in \mathbb{N}$, then the
hierarchies (\ref{2.12}) boil down to hierarchies (\ref{2.5}) and the
Lax--Novikov equations (\ref{2.11}) reduce to the commutativity conditions (%
\ref{2.4}). In particular, if the bracket $[\cdot ,\cdot ]$ is such that
equations (\ref{2.12}) give rise to integrable systems in $d$ independent
variables, then equations (\ref{2.5}) yield integrable systems in $d-1$
independent variables.

A standard construction of a commutative subalgebra spanned by $L_{i}$ whose
existence ensures commutativity of the flows (\ref{2.5}) and (\ref{2.12})
is, in the case of Lie algebras which admit an additional associative
multiplication $\circ $ which obeys the Leibniz rule
\begin{equation}
\lbrack a,b\circ c]=[a,b]\circ c+b\circ \lbrack a,c],  \label{2.1}
\end{equation}%
as follows: the commutative subalgebra is generated by rational powers of a
given element $L\in \mathfrak{g}$, cf.\ e.g.\ \cite{se,bsz2} and references
therein. This is also our case for (1+1)D and (2+1)D dispersionless systems,
when the Lie algebra in question is a Poisson algebra.\looseness=-1

However, in our (3+1)D setting, when the Leibniz rule is no longer required
to hold, this construction does not work anymore. In particular, it is the
case of (3+1)D dispersionless systems when the Lie algebra under study is a
Jacobi algebra. In order to circumvent this difficulty, instead of an
explicit construction of commuting $L_{i}$, we will \emph{impose} the
zero-curvature constraints (\ref{2.7}) on chosen elements $L_{i}\in
\mathfrak{g}$, $i\in \mathbb{N}$; it is readily seen that in the case of the
Jacobi algebra that we are interested in this can be done in a consistent
fashion.\label{scb}

Let us come back to the systems considered in the previous sections. For the
(3+1)D case consider a commutative and associative algebra $A$ of formal
series in $p$
\begin{equation}
A\ni f=\sum_{i}u_{i}p^{i}  \label{3.1}
\end{equation}%
with ordinary dot multiplication%
\begin{equation}
f_{1}\cdot f_{2}\equiv f_{1}f_{2},\qquad f_{1},f_{2}\in A.  \label{3.2}
\end{equation}%
The coefficients $u_{i}$ of these series are assumed to be smooth functions
of $x,y,z$ and infinitely many times $t_{1},t_{2},\dots$.

The Jacobi structure on $A$ will be induced by the contact bracket (\ref%
{con7})
\begin{equation}
\lbrack f_{1},f_{2}]\equiv \{f_{1},f_{2}\}_{C}=\displaystyle\frac{\partial
f_{1}}{\partial p}\frac{\partial f_{2}}{\partial x}-p\frac{\partial f_{1}}{%
\partial p}\frac{\partial f_{2}}{\partial z}+f_{1}\frac{\partial f_{2}}{%
\partial z}-(f_{1}\leftrightarrow f_{2}).  \label{cb}
\end{equation}%
Notice that this bracket is independent of $y$. As the unit element $e=1$
does not belong to the center of the Jacobi algebra, the Leibniz rule (\ref%
{2.1}) does not hold anymore, and instead we have
\begin{equation}
\{f_{1}f_{2},f_{3}\}_{C}=\{f_{1},f_{3}\}_{C}f_{2}+f_{1}\{f_{2},f_{3}%
\}_{C}-f_{1}f_{2}\{1,f_{3}\}_{C}.  \label{jlr}
\end{equation}

For (2+1)D and (1+1)D cases, if we drop the dependence on $z$ or on $z$ and $%
y$, this bracket reduces to the canonical Poisson bracket (\ref{hx2}) in one
degree of freedom
\begin{equation}
\{f_{1},f_{2}\}_{P}=\frac{\partial f_{1}}{\partial p}\frac{\partial f_{2}}{%
\partial x}-\frac{\partial f_{2}}{\partial p}\frac{\partial f_{1}}{\partial x%
}  \label{p0}
\end{equation}%
and the Jacobi algebra $\mathfrak{g}=(A,\cdot ,\{,\}_{C})$ reduces to the
Poisson algebra $\mathfrak{g}=(A,\cdot ,\{,\}_{P})$ respectively.

As for the choice of the splitting of the Jacobi algebra $\mathfrak{g}%
=(A,\cdot ,\{,\}_{C})$ into Lie subalgebras $\mathfrak{g}_{\pm }$ with $\Pi_{\pm }$ being projections onto the respective subalgebras, so that $\mathfrak{g}_{\pm }=\Pi_{\pm }(\mathfrak{g})$, it is readily checked that we
have two natural choices when the $R$'s defined by (\ref{2.14}) satisfy the
classical modified Yang--Baxter equation (\ref{2.3}) and thus are $R$%
-matrices. These two choices are of the form\looseness=-1
\begin{equation}
\Pi_{+}=\Pi_{\geqslant k},  \label{pro}
\end{equation}%
where $k=0$ or $k=1$, and by definition
\begin{equation*}
\Pi_{\geqslant k}\left( \sum\limits_{j=-\infty }^{\infty }a_{j}p^{j}\right)
=\sum\limits_{j=k}^{\infty }a_{j}p^{j}.
\end{equation*}%
Note that for (1+1)D and (2+1)D systems, associated with the Poisson algebra
$\mathfrak{g}=(A,\cdot ,\{,\}_{P}),$ the additional choice of $k=2$ in (\ref%
{pro}) is also admissible \cite{bsz1,bsz}.

\subsection{Integrable (3+1)D infinite-component hierarchies and their
lower-dimensional reductions}

\label{(3+1)}

We begin with the case of $k=0$ and the $n$th order Lax function from $A$ of
the form
\begin{equation}
L\equiv L_{n}=u_{n}p^{n}+u_{n-1}p^{n-1}+\cdots +u_{0}+u_{-1}p^{-1}+\cdots
,\qquad n>0  \label{5.1}
\end{equation}%
and let
\begin{equation}
B_{m}\equiv \Pi_{+}L_{m}=v_{m,m}p^{m}+v_{m,m-1}p^{m-1}+\cdots +v_{m,0},\qquad
m>0  \label{5.2}
\end{equation}%
where $u_{i}=u_{i}(x,y,z,\vec{t})$, $v_{m,j}=v_{m,j}(x,y,z,\vec{t})$, and $%
\vec{t}=(t_{1},t_{2},\dots )$.

Substituting $L$ and $B_{m}$ into the zero-curvature-type equations
\begin{equation}
L_{t_{m}}=\{B_{m},L\}_{C}+(B_{m})_{y}  \label{5.3}
\end{equation}%
we see that one can impose a natural constraint: $u_{n}=c_{n}$, $v_{m,m}=c_{m,m}$, where $c_{n},c_{m,m}\in \mathbb{R}$.

Then, if we put $c_{n}=c_{m,m}=1$, we get
\begin{equation}
L=p^{n}+u_{n-1}p^{n-1}+\cdots +u_{0}+u_{-1}p^{-1}+\cdots ,\quad n>0,
\label{5.6a}
\end{equation}%
\begin{equation}
B_{m}\equiv \Pi_{+}L_{m}=p^{m}+v_{m,m-1}p^{m-1}+\cdots +v_{m,0},\qquad m>0,
\label{5.6b}
\end{equation}%
and equations (\ref{5.3}) take the form
\begin{align}
0& =X_{r}^{m}[u,v_{m}],\qquad n<r<n+m,  \notag \\
(u_{r})_{t_{m}}& =X_{r}^{m}[u,v_{m}],\qquad r\leq n,\quad r\neq 0,\dots ,m-1,
\label{5.6cc} \\
(u_{r})_{t_{m}}& =X_{r}^{m}[u,v_{m}]+(v_{m,r})_{y},\qquad r=0,\dots ,m-1,
\notag
\end{align}%
where $v_{m}=(v_{m,0},\dots,v_{m,m}=1)$ and
\begin{equation}
\begin{array}{rcl}
X_{r}^{m}[u,v_{m}] & = & \displaystyle\sum%
\limits_{s=0}^{m}[sv_{m,s}(u_{r-s+1})_{x}-(r-s+1)u_{r-s+1}(v_{m,s})_{x} \\%
[5mm]
&  & \quad -(s-1)v_{m,s}(u_{r-s})_{z}+(r-s-1)u_{r-s}(v_{m,s})_{z}],%
\end{array}
\label{5.5}
\end{equation}%
where $u_{n}=1$ and $u_{r}=0$ for $r>n$. The fields $u_{r}$ for $r\leq n$
are dynamical variables while equations for $n+m>r>n$ can be seen as
nonlocal constraints on $u_{r}$ which define the variables $v_{m,s}$.
Observe that the additional dependent variables $v_{m,s}$ for different $m$
are by construction related to each other through the zero-curvature
equations (\ref{2.16}).

There is one more constraint in the Lax pair (\ref{5.1}) and (\ref{5.2}). The
first equation from the system (\ref{5.6cc}), i.e., the one for $r=n+m-1$,
takes the form
\begin{equation*}
(n-1)(v_{m,m-1})_{z}-(m-1)(u_{n-1})_{z}=0,
\end{equation*}%
so the system under study for $n>1$ admits a further constraint
\begin{equation}
v_{m,m-1}=\frac{(m-1)}{(n-1)}u_{n-1}.  \label{5.6d}
\end{equation}%
Thus, the final Lax pair takes the form
\begin{equation}
L=p^{n}+u_{n-1}p^{n-1}+\cdots +u_{0}+u_{-1}p^{-1}+\cdots,\quad n>0,
\label{5.6aa}
\end{equation}%
\begin{equation}
B_{m}=p^{m}+\tfrac{(m-1)}{(n-1)}u_{n-1}p^{m-1}+\cdots +v_{m,0},\qquad m>0
\label{5.6bb}
\end{equation}

It is readily seen that for $n=1$ the constraint (\ref{5.6d}) should be
replaced by $u_{0}=\mathrm{const}$. Let us consider this case in more
detail. Upon taking $u_{0}=0$, consider the Lax equation (\ref{5.3}) for
\begin{equation}
L=p+u_{-1}p^{-1}+u_{-2}p^{-2}+\cdots,  \label{5.7a}
\end{equation}%
\begin{equation}
B_{m}=p^{m}+v_{m,m-1}p^{m-1}+\cdots+v_{m,1}p+v_{m,0},\quad m>0;  \label{5.7aa}
\end{equation}
then the related system reads
\begin{align}
0& =(v_{m,r})_{y}+X_{r}^{m}[u,v_{m}],\qquad r=0,\dots,m-1,  \notag \\
(u_{r})_{t_{m}}& =X_{r}^{m}[u,v_{m}],\qquad r<0.  \label{5.7b}
\end{align}%
Thus, the simplest nontrivial case is $m=2$, so
\begin{equation*}
\ B_{2}=p^{2}+v_{1}p+v_{0}
\end{equation*}
and generates the following infinite-component system \cite{bls}
\begin{equation}\label{5.7}
\begin{array}{rcl}
(v_{1})_{y}&=&(v_{1})_{x}+(u_{-1})_{z}, \\[2mm]
(v_{0})_{y}& =&(v_{0})_{x}+(u_{-2})_{z}-2(u_{-1})_{x}+2u_{-1}(v_{1})_{z},
\\[2mm]
(u_{r})_{t_{2}}&
=&2(u_{r-1})_{x}-(u_{r-2})_{z}-(r+1)u_{r+1}(v_{0})_{x}+v_{0}(u_{r})_{z}
\\[2mm]
&&
+(r-1)u_{r}(v_{0})_{z}+v_{1}(u_{r})_{x}-ru_{r}(v_{1})_{x}+(r-2)u_{r-1}(v_{1})_{z},
\end{array}
\end{equation}
where $r<0$ and $v_{2,s}\equiv v_{s}$,  $s=0,1$.

We have a natural (2+1)D reduction of (\ref{5.7}) when $u_{r},v_{0}$ and $%
v_{1}$ are independent of $y$,
\begin{equation}
\begin{array}{rcl}
0 & = & (v_{1})_{x}+(u_{-1})_{z}, \\
0 & = & (v_{0})_{x}+(u_{-2})_{z}-2(u_{-1})_{x}+2u_{-1}(v_{1})_{z}, \\
(u_{r})_{t_{2}} & = &
2(u_{r-1})_{x}-(u_{r-2})_{z}-(r+1)u_{r+1}(v_{0})_{x}+v_{0}(u_{r})_{z} \\
&  &
+(r-1)u_{r}(v_{0})_{z}+v_{1}(u_{r})_{x}-ru_{r}(v_{1})_{x}+(r-2)u_{r-1}(v_{1})_{z},%
\end{array}
\label{5.8}
\end{equation}%
another (2+1)D reduction
\begin{align}
(v_{1})_{y}& =(u_{-1})_{z},  \notag \\
(v_{0})_{y}& =(u_{-2})_{z}+2u_{-1}(v_{1})_{z},  \label{5.8a} \\
(u_{r})_{t_{2}}&
=-(u_{r-2})_{z}+v_{0}(u_{r})_{z}+(r-1)u_{r}(v_{0})_{z}+(r-2)u_{r-1}(v_{1})_{z},
\notag
\end{align}%
when $u_{r},v_{0}$ and $v_{1}$ are independent of $x$, and yet another
(2+1)D reduction
\begin{align}
(v_{1})_{y}& =(v_{1})_{x},  \notag \\
(v_{0})_{y}& =(v_{0})_{x}-2(u_{-1})_{x},  \label{5.9} \\
(u_{r})_{t_{2}}&
=2(u_{r-1})_{x}-(r+1)u_{r+1}(v_{0})_{x}+v_{1}(u_{r})_{x}-ru_{r}(v_{1})_{x},
\notag
\end{align}%
when $u_{r},v_{0}$ and $v_{1}$ are independent of $z$.

Moreover, system (\ref{5.9}) admits a further reduction $v_{1}=0$ to the
form
\begin{align}
(v_{0})_{y}& =(v_{0})_{x}-2(u_{-1})_{x},  \label{5.9a} \\
(u_{r})_{t_{2}}& =2(u_{r-1})_{x}-(r+1)u_{r+1}(v_{0})_{x}+v_{1}(u_{r})_{x}.
\notag
\end{align}%
The system (\ref{5.9a}) reduces to $(1+1)$-dimensional system
\begin{equation}
(u_{r})_{t_{2}}=2(u_{r-1})_{x}-2(r+1)u_{r+1}(u_{-1})_{x},\quad r<0,
\label{5.10}
\end{equation}%
when $u_{i}$ are independent of $y$ and we put $v_{0}=2u_{-1}$.

Notice that (\ref{5.10}) is the well-known (1+1)D Benney system a.k.a.\ the Benney momentum chain \cite{be,km1}. From that point of view, the systems (\ref{5.9a}) and (\ref{5.8}) can be seen as natural (2+1)D extensions of the Benney
chain while the system (\ref{5.7}) represents a (3+1)D extension of the Benney system.

On the other hand, system (\ref{5.8a}) admits no reductions to $(1+1)$%
-dimensional systems. Note that for systems (\ref{5.7})--(\ref{5.10}) there
are no obvious finite-component reductions.

%Just as above,
For systems (\ref{5.1}), (\ref{5.2}) and (\ref{5.6a}), (\ref{5.6b}) we have $%
(2+1)$-dimensional and $(1+1)$-dimensional reductions of the same types as
above.

Now pass to the case of $k=1$, when $\Pi_{+}=\Pi_{\geqslant 1}$, and
consider the general case when
\begin{equation}
\begin{array}{rcl}
L & = & u_{n}p^{n}+u_{n-1}p^{n-1}+\cdots +u_{0}+u_{-1}p^{-1}+\dots ,\quad
n>0, \\[2mm]
B_{m} & = & v_{m,m}p^{m}+v_{m,m-1}p^{m-1}+\cdots +v_{m,1}p,\quad m>0,%
\end{array}
\label{5.27}
\end{equation}
from which we again obtain the hierarchies of infinite-component systems
\begin{equation}
\begin{array}{rcl}
0 & = & X_{r}^{m}[u,v_{m}],\qquad n<r\leq n+m, \\[2mm]
(u_{r})_{t_{m}} & = & X_{r}^{m}[u,v_{m}],\qquad r\leq n,\quad r\neq
1,\dots,m, \\[2mm]
(u_{r})_{t_{m}} & = & X_{r}^{m}[u,v_{m}]+(v_{m,r})_{y},\qquad r=1,\dots,m,%
\end{array}
\label{5.28a}
\end{equation}%
where $v_{m}=(v_{m,1},\dots ,v_{m,m})$ and
\begin{equation}
\begin{array}{rcl}
X_{r}^{m}[u,v_{m}] & = & \displaystyle\sum%
\limits_{s=1}^{m}[sv_{m,s}(u_{r-s+1})_{x}-(r-s+1)u_{r-s+1}(v_{m,s})_{x} \\%
[5mm]
&  & \quad -(s-1)v_{m,s}(u_{r-s})_{z}+(r-s-1)u_{r-s}(v_{m,s})_{z}].%
\end{array}
\label{5.28b}
\end{equation}

For $n>1,m>1$ there is an additional constraint imposed on the Lax pair (\ref%
{5.27}). The first equation from the system (\ref{5.28b}), i.e., the one for
$r=n+m$, takes the form
\begin{equation*}
(n-1)u_{n}(v_{m,m})_{z}-(m-1)v_{m,m}(u_{n})_{z}=0,
\end{equation*}%
and hence, for $n>1,m>1$, admits the constraint
\begin{equation}
v_{m,m}=(u_{n})^{\frac{m-1}{n-1}}.  \label{5.6}
\end{equation}%
So, the final Lax pair takes the form
\begin{equation}
\begin{array}{rcl}
L & = & u_{n}p^{n}+u_{n-1}p^{n-1}+\cdots +u_{0}+u_{-1}p^{-1}+\cdots ,\quad
n>1, \\[2mm]
B_{m} & = & (u_{n})^{\frac{m-1}{n-1}}p^{m}+v_{m,m-1}p^{m-1}+\cdots
+v_{m,1}p,\qquad m>0.%
\end{array}
\label{5.66}
\end{equation}

For $n=1$ the constraint in question is replaced by $u_{1}=\mathrm{const}$.
Thus, consider again in detail the simplest case when $n=1$ and $u_{1}=1$
\begin{equation}
\begin{array}{rcl}
L & = & p+u_{0}+u_{-1}p^{-1}+\cdots, \\[2mm]
B_{m} & = & v_{m,m-1}p^{m}+v_{m,m-2}p^{m-1}+\dots +v_{m,1}p,\qquad m>0,%
\end{array}
\label{5.17}
\end{equation}%
when the associated system reads
\begin{equation}
\begin{array}{rcl}
0 & = & (v_{m,r})_{y}+X_{r}^{m}[u,v_{m}],\qquad r=1,\dots,m, \\[2mm]
(u_{r})_{t_{m}} & = & X_{r}^{m}[u,v_{m}],\qquad r<0. \label{5.18a}%
\end{array}%
\end{equation}

Thus, the simplest nontrivial case is $m=2$, so %\begin{equation*}
$B_{2}=v_{2}p^{2}+v_{1}p$ %\end{equation*}%
generates the following infinite-component system \cite{bls}
\begin{equation}
\begin{array}{rcl}
(v_{2})_{y} & = & (v_{2})_{x}+u_{0}(v_{2})_{z}+v_{2}(u_{0})_{z}, \\
(v_{1})_{y} & = &
(v_{1})_{x}+u_{0}(v_{1})_{z}+v_{2}(u_{-1})_{z}+2u_{-1}(v_{2})_{z}-2v_{2}(u_{0})_{x},
\\
(u_{r})_{t_{2}} & = &
v_{1}(u_{r})_{x}-ru_{r}(v_{1})_{x}+(r-2)u_{r-1}(v_{1})_{z}+2v_{2}(u_{r-1})_{x}
\\
&  & \quad
-(r-1)u_{r-1}(v_{2})_{x}-v_{2}(u_{r-2})_{z}+(r-3)u_{r-2}(v_{2})_{z},%
\end{array}
\label{5.22}
\end{equation}
where $r<1$ and $v_{2,s}\equiv v_{s}$, $s=1,2$.

We have a natural $(2+1)$-dimensional reduction of (\ref{5.22}) when $%
u_{r},v_{1}$ and $v_{2}$ are independent of $y$,
\begin{equation}
\begin{array}{rcl}
0 & = & (v_{2})_{x}+u_{0}(v_{2})_{z}+v_{2}(u_{0})_{z}, \\
0 & = &
(v_{1})_{x}+u_{0}(v_{1})_{z}+v_{2}(u_{-1})_{z}+2u_{-1}(v_{2})_{z}-2v_{2}(u_{0})_{x},
\\
(u_{r})_{t_{2}} & = &
v_{1}(u_{r})_{x}-ru_{r}(v_{1})_{x}+(r-2)u_{r-1}(v_{1})_{z}+2v_{2}(u_{r-1})_{x}
\\
&  & -(r-1)u_{r-1}(v_{2})_{x}-v_{2}(u_{r-2})_{z}+(r-3)u_{r-2}(v_{2})_{z}.%
\end{array}
\label{22.a}
\end{equation}%
On the other hand, if $u_{r},v_{1}$ and $v_{2}$ are independent of $x$, we
obtain from (\ref{5.22}) another $(2+1)$-dimensional system
\begin{align}
(v_{2})_{y}& =u_{0}(v_{2})_{z}+v_{2}(u_{0})_{z},  \notag \\
(v_{1})_{y}& =u_{0}(v_{1})_{z}+v_{2}(u_{-1})_{z}+2u_{-1}(v_{2})_{z},
\label{22b} \\
(u_{r})_{t_{2}}&
=(r-2)u_{r-1}(v_{1})_{z}-v_{2}(u_{r-2})_{z}+(r-3)u_{r-2}(v_{2})_{z}.  \notag
\end{align}%
Next, if $u_{r},v_{1}$ and $v_{2}$ in (\ref{5.22}) are independent of $z$, we arrive at the third $(2+1)$-dimensional system
\begin{align}
(v_{2})_{y}& =(v_{2})_{x},  \notag \\
(v_{1})_{y}& =(v_{1})_{x}-2v_{2}(u_{0})_{x},  \label{5.22d} \\
(u_{r})_{t_{2}}&
=v_{1}(u_{r})_{x}-ru_{r}(v_{1})_{x}+2v_{2}(u_{r-1})_{x}-(r-1)u_{r-1}(v_{2})_{x},
\notag
\end{align}%
whence, after the substitution $v_{2}=\mathrm{const}=1$, we obtain
\begin{align}
(v_{1})_{y}& =(v_{1})_{x}-2(u_{0})_{x},  \label{22c} \\
(u_{r})_{t_{2}}& =v_{1}(u_{r})_{x}-ru_{r}(v_{1})_{x}+2(u_{r-1})_{x}.  \notag
\end{align}%
If $u_{r},v_{1}$ and $v_{2}$ are independent of both $y$ and $z$, we can put
$v_{1}=2u_{0}$ and obtain
\begin{equation}
(u_{r})_{t_{2}}=2(u_{r-1})_{x}+2u_{0}(u_{r})_{x}-2ru_{r}(u_{0})_{x}.
\label{22d}
\end{equation}%
Finally, when $u_{r},v_{1}$ and $v_{2}$ are independent of both $y$ and $x$,
we have
\begin{equation}
(u_{r})_{t_{2}}=(r-2)u_{r-1}(v_{1})_{z}-v_{2}(u_{r-2})_{z}+(r-3)u_{r-2}(v_{2})_{z},
\label{22e}
\end{equation}%
where a reduction
\begin{equation*}
v_{2}=au_{0}^{-1},\quad v_{1}=-au_{-1}u_{0}^{-2},
\end{equation*}%
was performed, and $a\in \mathbb{R}$ is an arbitrary constant. Thus, in this
case the system under study is rational (rather than polynomial) in $u_{0}$.

\subsection{Finite-component reductions}

\label{fc} For $k=0$, in contrast with the simplest case (\ref{5.7a}), we do
have natural reductions to finite-component systems. Namely, they are of
the form%
\begin{equation}
\begin{array}{rcl}
L & = & u_{n}p^{n}+u_{n-1}p^{n-1}+\cdots +u_{r}p^{r},\quad r=0,1, \\[2mm]
B_{m} & = & (u_{n})^{\frac{m-1}{n-1}}p^{m}+v_{m,m-1}p^{m-1}+\cdots +v_{m,0},%
\end{array}
\label{5.14}
\end{equation}%
and
\begin{equation}
\begin{array}{rcl}
L & = & p^{n}+u_{n-1}p^{n-1}+\cdots +u_{r}p^{r},\quad r=0,1, \\
B_{m} & = & p^{m}+\displaystyle\tfrac{(m-1)}{(n-1)}u_{n-1}p^{m-1}+\cdots
+v_{m,0}.%
\end{array}
\label{5.14a}
\end{equation}
The case (\ref{5.14a}) for $r=0$ was considered for the first time in \cite{aslmp} while the remaining cases was analyzed in \cite{bls}. Notice that in
(\ref{5.14}) and (\ref{5.14a}) for $r=0$ we have $L=B_{n}$, and hence the
variable $y$ can be identified with $t_{n}$. Then equations (\ref{5.3})
coincide with the zero-curvature equations (\ref{2.16}), and the Lax--Novikov
equation (\ref{2.11}) reduces to equation (\ref{2.15}).

Another class of natural reductions to finite-component systems arises for $%
k=1$ \cite{bls}. Indeed, for $n>1$ we have
\begin{equation}
\begin{array}{rcl}
L & = & u_{n}p^{n}+u_{n-1}p^{n-1}+\cdots +u_{r}p^{r},\quad r=1,0,-1,\dots \\%
[2mm]
B_{m} & = & (u_{n})^{\tfrac{m-1}{n-1}}p^{m}+v_{m,m-1}p^{m-1}+\cdots +v_{m,1}p%
\end{array}
\label{22a}
\end{equation}%
while for $n=1$%
\begin{equation}
\begin{array}{rcl}
L & = & p+u_{0}+u_{-1}p^{-1}+\cdots +u_{r}p^{r},\quad r=0,1,-1,\dots \\
B_{m} & = & v_{m,m}p^{m}+v_{m,m-1}p^{m-1}+\dots +v_{m,1}p,\qquad m>1.%
\end{array}
\label{5.23}
\end{equation}

In closing we point out a large class of finite-component reductions of the
hierarchy associated with (\ref{5.6a}) and (\ref{5.6b}) for $k=0$. The
reductions in question for $L$ (\ref{5.6a}) are given by rational Lax
functions, cf.\ \cite{kri2, bsz3} and references therein for the (1+1)D
case, namely,
\begin{equation}  \label{rat}
\begin{array}{rcl}
L & = & p^n + \displaystyle\sum_{j=0}^{n-1}u_j p^j + \sum\limits_{i=1}^{k}%
\frac{a_{i}}{(p - r_i)},\qquad n>1, \quad k>0,%
\end{array}%
\end{equation}
where $u_j$, $a_i$ and $r_i$ are unknown functions; in this case $B_m$ are
still given by (\ref{5.6b}).
%$\boldsymbol{u}=(u_0,\dots,u_{n-1},a_1,\dots,a_k,r_1,\dots,r_k,v_0,\dots,v_{m-1},b_1,\dots,b_{\ell},s_1,\dots,s_{\ell})^T$.

\begin{example}
\emph{First let us begin with the case of $k=0$ and the simplest Lax pair
from (\ref{5.14a}) when $n=2$ and $m=3$}
\begin{equation}
L=p^{2}+u_{1}p+u_{0},\quad B=p^{3}+2u_{1}p^{2}+v_{1}p+v_{0},  \label{kp1}
\end{equation}%
\emph{The zero-curvature-type Lax equation}
\begin{equation*}
L_{t}=\{B,L\}_{C}+(B)_{y}
\end{equation*}%
\emph{generates a four-component system \cite{aslmp}}
\begin{equation}
\hspace*{-5mm}
\begin{array}{rcl}
(u_{0})_{t} & \!=\! &
(v_{0})_{y}+v_{0}(u_{0})_{z}+v_{1}(u_{1})_{x}-u_{0}(v_{0})_{z}-u_{1}(v_{0})_{x},
\\
(u_{1})_{t} & \!=\! &
(v_{1})_{y}-2(v_{0})_{x}+4u_{1}(u_{0})_{x}-u_{1}(v_{1})_{x}+v_{1}(u_{1})_{x}+v_{0}(u_{1})_{z}-u_{0}(v_{1})_{z},
\\
0 & \!=\! & (v_{1})_{z}-(u_{1})_{x}-2(u_{0})_{z}-2u_{1}(u_{1})_{z}, \\
0 & \!=\! &
(v_{0})_{z}+3(u_{0})_{x}+2(u_{1})_{y}-2(v_{1})_{x}+2u_{1}(u_{1})_{x}-2u_{1}(u_{0})_{z}-2u_{0}(u_{1})_{z}.%
\end{array}
\hspace{-4mm}  \label{kp2}
\end{equation}
\emph{which is a natural (3+1)D extension of the (2+1)D dispersionless
Kadomtsev-Petviashvili (dKP) equation.}
\emph{Indeed, upon assuming that all fields
are independent of $z$, and that $u_{1}=0$ and $v_{1}=\frac{3}{2}u_{0}$, and
denoting $u_{0}\equiv u$ and $v_0\equiv v$, the system (\ref{kp2}) reduces
to the form}
\begin{equation}
u_{t}=v_{y}+\tfrac{3}{2}u\,u_{x},\ \ 3u_{y}=4v_{x}\Longrightarrow (u_{t}-%
\tfrac{3}{2}u\,u_{x})_{x}=\tfrac{3}{4}u_{yy}  \label{kp4}
\end{equation}%
\emph{where the last equation in (\ref{kp4}) is, up to a suitable rescaling
of independent variables, nothing but the celebrated dKP equation, also
known as the three-dimensional Khokhlov--Zabolotskaya \cite{kz} equation.}
\end{example}

\begin{example}
\emph{Consider again our system (\ref{xx}), which is a particular case of (%
\ref{5.23}) for $k=1$ and $r=-1$, with notation $u_{-1}\equiv u_{1}$, being
the first member of the hierarchy (\ref{5.23}) generated by the Lax functions}
\begin{equation*}
L=p+u_{0}+u_{1}p^{-1},\quad B_{2}=v_{2}p^{2}+v_{1}p,
\end{equation*}%
\emph{and takes the known form (\ref{3dd})}%
\begin{equation}
\begin{array}{rcl}
(u_{1})_{t_{2}} & = & u_{1}(v_{1})_{x}+v_{1}(u_{1})_{x}, \\
(u_{0})_{t_{2}} & = &
-2u_{1}(v_{1})_{z}+v_{1}(u_{0})_{x}+u_{1}(v_{2})_{x}+2v_{2}(u_{1})_{x}, \\
(v_{1})_{y} & = &
(v_{1})_{x}+2u_{1}(v_{2})_{z}+v_{2}(u_{1})_{z}+u_{0}(v_{1})_{z}-2v_{2}(u_{0})_{x},
\\
(v_{2})_{y} & = & (v_{2})_{x}+u_{0}(v_{2})_{z}+v_{2}(u_{0})_{z}.%
\end{array}
\label{ut2}
\end{equation}%
\emph{The second member of the hierarchy is generated by}
\begin{equation*}
L=p+u_{0}+u_{1}p^{-1},\quad B_{3}=w_{3}p^{3}+w_{2}p^{2}+w_{1}p
\end{equation*}%
\emph{and has the form}
\begin{equation}
\begin{array}{rcl}
(u_{1})_{t_{3}} & = & u_{1}(w_{1})_{x}+w_{1}(u_{1})_{x}, \\
(u_{0})_{t_{3}} & = &
w_{1}(u_{0})_{x}-2u_{1}(w_{1})_{z}+u_{1}(w_{2})_{x}+2w_{2}(u_{1})_{x}, \\
(w_{1})_{y} & = &
(w_{1})_{x}+w_{2}(u_{1})_{z}-u_{1}(w_{3})_{x}-2w_{2}(u_{0})_{x}+2u_{1}(w_{2})_{z}
\\
&  & +u_{0}(w_{1})_{z}-3w_{3}(u_{1})_{x}, \\
(w_{2})_{y} & = &
(w_{2})_{x}-3w_{3}(u_{0})_{x}+2w_{3}(u_{1})_{z}+w_{2}(u_{0})_{z} \\
&  & +u_{0}(w_{2})_{z}+2u_{1}(w_{3})_{z}, \\
(w_{3})_{y} & = & (w_{3})_{x}+u_{0}(w_{3})_{z}+2w_{3}(u_{0})_{z},%
\end{array}
\label{ut3}
\end{equation}%
\emph{Commutativity of the flows associated with $t_{2}$ and $t_{3}$, i.e.}
\begin{equation*}
\left( (u_{i})_{t_{2}}\right) _{t_{3}}=\left( (u_{i})_{t_{3}}\right)
_{t_{2}},\quad i=0,1,
\end{equation*}%
\emph{can be checked using the set of relations}%
\begin{equation*}
\begin{array}{rcl}
(v_{1})_{z} & = & -\displaystyle\frac{v_{2}}{w_{3}}(w_{3})_{x}-\frac{%
v_{2}w_{2}}{4w_{3}^{2}}(w_{3})_{z}+\frac{v_{2}}{2w_{3}}(w_{2})_{z}+\frac{3}{2%
}(v_{2})_{x},\quad (v_{2})_{z}=\displaystyle\frac{v_{2}}{2w_{3}}(w_{3})_{z},
\\[3mm]
(w_{1})_{t_{2}} & = & v_{1}(w_{1})_{x}-w_{1}(v_{1})_{x}+(v_{1})_{t_{3}}, \\%
[2mm]
(w_{2})_{t_{2}} & = &
v_{1}(w_{2})_{x}-w_{1}(v_{2})_{x}+2v_{2}(w_{1})_{x}-2w_{2}(v_{1})_{x}+(v_{2})_{t_{3}},
\\[2mm]
(w_{3})_{t_{2}} & = & \displaystyle\frac{v_{2}w_{2}}{2w_{3}}(w_{2})_{z}-%
\frac{w_{2}}{2}(v_{2})_{x}-\frac{v_{2}w_{2}^{2}}{4w_{3}^{2}}(w_{3})_{z}+%
\frac{(v_{1}w_{3}-v_{2}w_{2})}{w_{3}}(w_{3})_{x} \\[4mm]
&  & -v_{2}(w_{1})_{z}+2v_{2}(w_{2})_{x}-3w_{3}(v_{1})_{x},%
\end{array}%
\end{equation*}%
\emph{which is equivalent to the zero-curvature equation}
\begin{equation}
(B_{2})_{t_{3}}-(B_{3})_{t_{2}}+\{B_{2},B_{3}\}_{C}=0.  \label{zcr-t2t3}
\end{equation}%
\emph{Moreover, the compatibility conditions}
\begin{equation*}
\left( (v_{i})_{y}\right) _{z}=\left( (v_{i})_{z}\right) _{y},\quad i=1,2,
\end{equation*}%
\emph{are also satisfied by virtue of (\ref{ut2}) and (\ref{zcr-t2t3}).}

\emph{When }$u_{r}$ \emph{and} $v_{j}$ \emph{are independent of} $z$, \emph{we
obtain} $(2+1)D$ \emph{systems with additional constraints} $v_{2}$=$\mathrm{%
const\ } $$=\tfrac{1}{2},w_{3}=\mathrm{const}=\tfrac{1}{3}$%
\begin{equation}
\hspace*{-10mm}%
\begin{array}{rcl}
(u_{1})_{t_{2}} & = & u_{1}(v_{1})_{x}+v_{1}(u_{1})_{x}, \\
(u_{0})_{t_{2}} & = & v_{1}(u_{0})_{x}+(u_{1})_{x}, \\
(v_{1})_{y} & = & (v_{1})_{x}-(u_{0})_{x},%
\end{array}%
\ \
\begin{array}{rcl}
(u_{1})_{t_{3}} & = & u_{1}(w_{1})_{x}+w_{1}(u_{1})_{x}, \\
(u_{0})_{t_{3}} & = & w_{1}(u_{0})_{x}+u_{1}(w_{2})_{x}+2w_{2}(u_{1})_{x},
\\
(w_{1})_{y} & = & (w_{1})_{x}-(u_{1})_{x}-2w_{2}(u_{0})_{x}, \\
(w_{2})_{y} & = & (w_{2})_{x}-(u_{0})_{x}.%
\end{array}%
\hspace{-2mm}  \label{100}
\end{equation}%
\emph{see also (\ref{e8}).}

\emph{When $u_{r}$ and $v_{j}$ are independent of $x$, we obtain other $(2+1)D$ systems making use of a naturally arising extra constraint $u_{1}=\tfrac{1%
}{2}$, namely,}%
\begin{equation}
\begin{array}{rcl}
(u_{0})_{t_{2}} & = & -(v_{1})_{z}, \\
(v_{1})_{y} & = & (v_{2})_{z}+u_{0}(v_{1})_{z}, \\
(v_{2})_{y} & = & (u_{0}v_{2})_{z}%
\end{array}%
\quad
\begin{array}{rcl}
(u_{0})_{t_{3}} & = & -(w_{1})_{z}, \\
(w_{1})_{y} & = & (w_{2})_{z}+u_{0}(w_{1})_{z}, \\
(w_{2})_{y} & = & (w_{3})_{z}+(u_{0}w_{2})_{z}, \\
(w_{3})_{y} & = & u_{0}(w_{3})_{z}+2w_{3}(u_{0})_{z}.%
\end{array}
\label{102}
\end{equation}

\emph{Further reduction of (\ref{100}) and (\ref{102}) by assuming that $%
u_{r}$, $v_{j}$ and $w_{k}$ are independent of $y$ leads to $(1+1)$D systems
of the form}%
\begin{equation}
\begin{array}{rcl}
(u_{1})_{t_{2}} & = & (u_{1}u_{0})_{x}, \\[2mm]
(u_{0})_{t_{2}} & = & (u_{1}+\tfrac{1}{2}u_{0}^{2})_{x},%
\end{array}%
\quad
\begin{array}{rcl}
(u_{1})_{t_{3}} & = & (u_{1}u_{0}^{2}+u_{1}^{2})_{x}, \\[2mm]
(u_{0})_{t_{3}} & = & (\tfrac{1}{3}u_{0}^{3}+2u_{0}u_{1})_{x},%
\end{array}
\label{104}
\end{equation}%
\emph{where we put $v_{1}=u_{0}$, $w_{2}=u_{0}$ and} $w_{1}=u_{0}^{2}+u_{1}.$

\emph{Likewise, the reduction of (\ref{102}) and (\ref{104}) by assuming
that $u_{r}$, $v_{j}$ and $w_{k}$ are independent of $y$ leads to $(1+1)D$
systems of the form}
\begin{equation}
(u_{0})_{t_{2}}=\tfrac{1}{2}(u_{0}^{-2})_{z},\qquad (u_{0})_{t_{3}}=-\tfrac{3%
}{4}(u_{0}^{-4})_{z},  \label{106}
\end{equation}%
\emph{thanks to the relations}
\begin{equation*}
v_{2}=u_{0}^{-1},\quad v_{1}=-u_{0}^{-2},\quad w_{3}=u_{0}^{-2},\quad
w_{2}=-u_{0}^{-3},\quad w_{1}=\tfrac{3}{4}u_{0}^{-4}.
\end{equation*}
\end{example}

\begin{example}
\emph{
Consider the first member of the hierarchy (\ref{5.23}) for $r=1$,}
\emph{generated by the Lax pair}
\begin{equation*}
L=up^{3}+wp^{2}+vp,\quad B_{2}=u^{\frac{1}{2}}p^{2}+sp,
\end{equation*}%
\emph{and takes the form }%
\begin{equation}
\begin{array}{lll}
u_{t} & = & su_{x}-3us_{x}+ws_{z}+2u^{\frac{1}{2}}w_{x}-u^{\frac{1}{2}%
}v_{z}-wu^{-\frac{1}{2}}u_{x}, \\
w_{t} & = & sw_{x}-2ws_{x}+2u^{\frac{1}{2}}v_{x}-\frac{1}{2}vu^{-\frac{1}{2}%
}u_{x}+\frac{1}{2}u^{-\frac{1}{2}}u_{y}, \\
v_{t} & = & v_{x}+s_{y}-vs_{x}, \\
0 & = & 2us_{z}-u^{\frac{1}{2}}w_{z}+\frac{1}{2}u^{\frac{1}{2}}u_{x}+\frac{1%
}{2}wu^{\frac{1}{2}}u_{z},%
\end{array}
\label{107}
\end{equation}%
\emph{where we put }$\emph{u}_{3}=u,\ u_{2}=w,\ u_{1}=v,\ v_{2,1}=s$ \emph{%
and }$t_{2}=t.$ \emph{The }$(2+1)D$ \emph{reduction with all fields
independent of }$z$ \emph{and }$u=1,\ s=\frac{2}{3}w$ \emph{reads }%
\begin{equation}
\begin{array}{lll}
w_{t} & = & 2w_{x}-\frac{2}{3}ww_{x}, \\
v_{t} & = & v_{x}-\frac{2}{3}vw_{x}+\frac{2}{3}w_{y},%
\end{array}
\label{108}
\end{equation}%
\emph{while the case when all fields are independent of }$x$ \emph{takes the
form }%
\begin{equation}
\begin{array}{lll}
u_{t} & = & ws_{z}-u^{\frac{1}{2}}v_{z}, \\
w_{t} & = & (u^{\frac{1}{2}})_{y}, \\
v_{t} & = & s_{y}, \\
0 & = & 2us_{z}-u^{\frac{1}{2}}w_{z}+w(u^{\frac{1}{2}})_{z},%
\end{array}
\label{109}
\end{equation}%
\emph{or equivalently }%
\begin{equation}
\begin{array}{l}
2a_{t}a_{tt}+a_{t}b_{yz}-a_{y}b_{tz}=0, \\
2a_{t}^{2}b_{tz}+a_{y}a_{tz}-a_{t}a_{yz}=0,%
\end{array}
\label{110}
\end{equation}%
\emph{where }$a_{t}=u^{\frac{1}{2}},\ a_{y}=w,\ b_{t}=s,\ b_{y}=v.$ \emph{%
The $(1+1)D$ reductions of (\ref{108}) and (\ref{109}), when all fields are
additionally independent of} $y$, \emph{take the form}%
\begin{equation*}
\begin{array}{lll}
w_{t} & = & 2v_{x}-\frac{2}{3}ww_{x}, \\
v_{t} & = & v_{x}-\frac{2}{3}vw_{x},%
\end{array}%
\end{equation*}%
\emph{and }%
\begin{equation*}
u_{t}+u^{-\frac{3}{2}}u_{z}=0,
\end{equation*}%
\emph{where} $v=0$ \emph{end }$w=2$.\emph{\ }
\end{example}

\subsection{Acknowledgments}

The research of AS was supported in part by the Ministry of Education, Youth and Sport of the Czech Republic (M\v{S}MT \v{C}R) under RVO funding for I\v{C}%
47813059 and the Grant Agency of the Czech Republic (GA \v{C}R) under grant P201/12/G028.

AS gratefully acknowledges warm hospitality extended to him in the course of his visit to the Adam Mickiewicz University where a substantial part of the present paper was written.

A number of computations in the present paper were performed using the software \emph{Jets} for Maple \cite{jets} whose use is acknowledged with gratitude.

The authors are pleased to thank B.M. Szablikowski for helpful comments. AS also thanks A. Borowiec and R. Vitolo for stimulating discussions.

\end{document}